\documentclass[5p]{elsarticle}

\usepackage{graphicx}
\usepackage[bottom]{footmisc}
\usepackage[draft]{hyperref}
\usepackage{caption}
\usepackage{multirow}
\usepackage[graphicx]{realboxes}
\usepackage{footnote}
\usepackage{subcaption}
\usepackage{lscape}
\usepackage{natbib}
\bibpunct{(}{)}{;}{a}{}{,}
\newcommand{\inertiaunit}{$J s^{-1/2} K^{-1} m^{-2}$}
\usepackage{hyperref}
\usepackage{geometry}
\usepackage{color}
\usepackage{cuted, ragged2e}
\usepackage{lipsum, array, booktabs, caption}
\usepackage{multirow}
\usepackage{amsmath}
\usepackage{amsfonts}
\usepackage{amssymb}
\usepackage{aas_macros}
\begin{document}
\begin{frontmatter}
\newpage
\title{Constraining the Thermal Properties of Planetary Surfaces using Machine Learning: Application to Airless Bodies}

\author[1]{Saverio Cambioni\corref{cor1}}
\cortext[cor1]{cambioni@lpl.arizona.edu}
\author[2]{Marco Delbo}
\author[2]{Andrew J. Ryan}
\author[3]{Roberto Furfaro}
\author[1]{Erik Asphaug}

\address[1]{Lunar and Planetary Laboratory, 1629 East University Boulevard, Tucson, AZ, 85721-0092, USA}
\address[2]{Universit\'e C\^ote d'Azur, Observatoire de la C\^ote d'Azur, CNRS, Laboratoire Lagrange
BD de l'Observatoire CS 34229, 06304 Nice Cedex 4, France}
\address[3]{Systems and Industrial Engineering Department, University of Arizona, Tucson, AZ 85721}

\journal{Icarus}

\begin{abstract}
We present a new method for the determination of the surface properties of airless bodies from measurements of the emitted infrared flux. Our approach uses machine learning techniques to train, validate, and test a neural network representation of the thermophysical behavior of the atmosphereless body given shape model, illumination and observational geometry of the remote sensors. The networks are trained on a dataset of thermal simulations of the emitted infrared flux for different values of surface rock abundance, roughness, and values of the thermal inertia of the regolith and of the rock components. These surrogate models are then employed to retrieve the surface thermal properties by Markov Chain Monte Carlo Bayesian inversion of observed infrared fluxes. We apply the method to the inversion of simulated infrared fluxes of asteroid (101195) Bennu -- according to a geometry of observations similar to those planned for NASA's OSIRIS-REx mission -- and infrared observations of asteroid (25143) Itokawa.  In both cases, the surface properties of the asteroid -- such as surface roughness, thermal inertia of the regolith and rock component, and relative rock abundance -- are retrieved; the contribution from the regolith and rock components are well separated. For the case of Itokawa, we retrieve a rock abundance of about 85\% for pebbles larger than the diurnal skin depth, which is about 2 $cm$. The thermal inertia of the rock is found to be lower than the expected value for LL chondrites, indicating that the rocks on Itokawa could be fractured.  The average thermal inertia of the surface is around 750 \inertiaunit~and the measurement of thermal inertia of the regolith corresponds to an average regolith particle diameter of about 10 $mm$, consistently with in situ measurements as well as results from previous studies.
\end{abstract}

\begin{keyword}
Asteroids, surfaces \sep Regoliths \sep Infrared observations \sep Machine Learning 

\end{keyword}

\end{frontmatter}
\section{Introduction}
\label{intro}
Surfaces of asteroids carry information about the composition, geology, formation and evolution of these bodies
\citep[][and references therein]{Murdoch2015aste.book..767M}. All the asteroids observed so far have been found to be covered in regolith, which is a layer of fragmented and unconsolidated material \citep[e.g.,][]{1991McKay} on an airless world. Small asteroids typically display substantial regolith with a variety of features. NASA's NEAR-Shoemaker mission to asteroid (433) Eros imaged debris aprons, fine-grained ``ponded" deposits, talus cones, and bright and dark streamers on steep slopes indicative of efficient downslope movement of the regolith \citep{2002Robinson}.  JAXA's Hayabusa mission found a considerable amount of regolith distributed nonuniformly on the surface of the small ($\sim$300-m diameter) near-Earth asteroid (25143) Itokawa \citep{2007Miyamoto}, which led to the question of how the regolith is segregated. In the case of larger airless bodies like the Moon, size-sorted regolith is locally concentrated, typically covering a bedrock. In the case of small bodies there is evidence of unconsolidated surface materials extending several hundred meters deep \citep{2002Robinson,Vincent2012P&SS...66...79V} and, in many cases, it extends throughout the asteroid, leading to the concept that they are rubble piles held together by gravity and cohesive forces \citep{2002aste.book_Rich}. These cohesive forces, including van der Waals interactions between the constituent grains, could play important roles in shaping the geomorphology of small rubble-pile asteroids \citep{2010Scheeres, 2014Rozitis} and their responses to collisions \citep{2015aste.book_Asphaug}. The presence of unconsolidated surface material, however, is not diagnostic of a rubble pile configuration; a thick layer of regolith has been recorded also on differentiated or partially differentiated bodies, such as (4) Vesta \citep[and references therein]{2002keil} and (21) Lutetia \citep{Vincent2012P&SS...66...79V,2012WeissLut}.

The properties of asteroid regoliths and in general the geomorphology of the surfaces of these bodies are diverse, and vary with their sizes and compositions. Only a few asteroids have been examined in close physical detail, by spacecraft, and a systematic exploration is being conducted by Earth-based telescopes and radars \citep[][and references therein]{Murdoch2015aste.book..767M}.
A few general conclusions, however, can already be drawn. It appears that large asteroids with sizes in the range of some hundreds $km$ are found to have very fine regolith grains of the order of tens to some hundreds of $\mu m$, while $km$-sized and smaller asteroids have coarser regoliths, whose typical grains can reach $cm$ or $dm$ sizes \citep{Vernazza2012Icar..221.1162V,2013Gundlach}. Itokawa regolith is gravel-sized in texture, while Eros regolith is dusty. This is not unexpected \citep{2011Hartzell}, since larger asteroids have more gravity and can hold onto more of their near-surface fragmentation products. 

Regolith evolves in time in response to processes such as impacts, thermal cycling and space weathering. Despite their low gravity, it is believed that most asteroids form from the accumulation of catastrophic disruption fragments \citep{Benz1999Icar..142....5B,Michel2013A&A...554L...1M} resulting in the creation of surfaces composed of boulders and smaller remnants. There is also indication that boulders are subsequently broken up into the smaller particles constituting the regolith by micrometeoroid impacts \citep{Horz1975Moon...13..235H,Horz1997M&PS...32..179H,Basilevsky2013P&SS...89..118B}, a fraction of them held by the asteroid and a fraction of them escaping the asteroid. Boulders are broken more gently by thermal fatigue cracking, caused by the numerous temperature variations between day and night \citep{Delbo2014Natur.508..233D,Molaro:2011tr,ElMir2016LPI....47.2586E}.

A realization of planetary sciences of the last $\sim$20 years is that the properties of the surfaces (and therefore of regolith) affect the orbital evolution of asteroids: the finite thermal inertia of the regolith causes the asteroid to emit more thermal photons in the afternoon compared to the morning emission. This imbalanced emission produces an acceleration that has a components along the orbital velocity of the asteroid, causing body's orbital semi-major axis to change with time, i.e., the so-called Yarkovsky effect \citep{Vokrouhlicky1998A&A...335.1093V,Bottke2006AREPS..34..157B,Vokrouhlicky2015aste.book..509V}. A related mechanism, known with the acronym of YORP effect, is related to the force torque due to reflection and emission of visibile and thermal photons from the asteroid surface. YORP produces a variation of the rotation period and direction of the spin axis.  Interestingly, the tangential component of the YORP effect is strongly affected by the amount of rocks with respect to that of fines \citep{Golubov2012ApJ...752L..11G}.

Knowledge of surface temperatures, determinations of thermal inertia, and the presence and abundance of visible rocks and their size distributions, are crucial in defining the thermal and mechanical environments on an asteroid's surface. These properties are used to estimate ``sampleability'' of a surface, to plan near-surface, landing, and sampling operations of active space missions, such as JAXA's Hayabusa 2 \citep{2012Abe} and NASA's OSIRIS-REx \citep{Lauretta2014M&PS..tmp..113L}, and, in the future, JAXA's MMX mission to Phobos, and human interactions with asteroids and the small moons of Mars. As an example, the sampling device of OSIRIS-REx, TAGSAM,  which is designed to collect and return at least 60g of the regolith of the asteroid (101955) Bennu, cannot collect grains larger than $\sim$2 cm, which illustrates how accurate estimates of grain size in asteroid regolith influence sampleability.\\

The present work focuses on the development of a new method to determine regolith and rock physical properties on airless bodies, coupling machine learning techniques and Bayesian inversion of observed infrared fluxes. In the following Section \ref{method} we introduce the state-of-the-art techniques and the proposed new methodology. In Section \ref{Bennu} we invert simulated thermal infrared observations by NASA's OSIRIS-REx. In Section \ref{Itokawa} we interpret thermal infrared observations of the near-Earth asteroid Itokawa to retrieve surface properties (namely: surface roughness, rock abundance, thermal inertia of the rock and regolith components). We conclude the paper with a discussion of the potentialities of the method and future applications.

\section{Methods}
\label{method}
\subsection{Infrared remote sensing of planetary surfaces}

A powerful technique to estimate the properties of regoliths on airless bodies consists in the analysis of the heat emitted in response to the changing diurnal insolation. This heat is typically measured by means of remote-sensing observations in the thermal infrared, which allows the determination of physical quantities such as the thermal inertia ($\Gamma=\sqrt{k \rho c}$, where $k$ is the thermal conductivity of the surface material, $\rho$ is the bulk density of the granular material, and $c$ is the mean specific heat), the degree of roughness, and the effective grain size of the regolith \citep{2013Gundlach,2015Delbo,Davidsson2015Icar..252....1D}.
In particular, surfaces with low thermal inertia respond to change of the illumination with instantaneous changes of their temperature, causing large temperature differences between day and night \citep[][and references therein]{Harris2002aste.conf..205H,2015Delbo}, whereas in the case of higher values of the thermal inertia, the surface responds more slowly to changes of the insolation, resulting in smoother diurnal temperature curves, with smaller differences in temperature between the day and the night. The granularity, porosity, degree of compaction, and lateral and vertical variations in these parameters within the field of view of the infrared instrument also affect the effective  thermal inertia of the regolith. Regolith temperatures are influenced by the presence and amount of large rocks, in addition to factors such as the thermophysical properties of the regolith fines, latitude and local slopes, and radiative heating from adjacent crater walls. 

The infrared radiance spectrum of a multi-material surface cannot be matched to a single blackbody at a single temperature but rather is a combination of Planck curves in radiance space. The infrared flux of the surface is contributed by the emission from both material (rock and regolith), weighted according to their abundance. It is common experience that rocks have higher thermal inertia of fines, such as sand. Rocks remain warmer than regolith during the night, due to their higher thermal inertia, so that mutual heating by exchange of radiation between warm and cold surfaces may increase regolith temperatures in rocky areas. Our definition of a rock is every boulder that is larger than the thermal skin depth of the surface, defined as a function of the average thermal conductivity $k$, bulk density $\rho$, specific heat $c$ of the surface layer, and the spin rotation period $P$:

\begin{equation}
\label{skin}
    \delta = \sqrt{\frac{k}{\rho c}}\sqrt{\frac{P}{\pi}} = \frac{\Gamma}{\rho c}\sqrt{\frac{P}{\pi}}
\end{equation}
\noindent
where $\Gamma$ is the thermal inertia of the surface material. Under this assumption, the infrared flux is the linear superposition of two spectra: the emission of fine regolith and the emission of solid rocks. The coefficient of the sum that multiplies each spectrum is the areal rock abundance (hereafter called rock abundance, or RA), for the rock and the regolith components, $RA$ and $(1-RA)$ respectively, i.e.,

\begin{equation}
\label{superpos}
    F(\lambda, t)=f_{rock}(\lambda,t) ~RA + f_{regolith}(\lambda,t)~ (1-RA)
\end{equation}
\noindent
where $\lambda$ is the wavelength and $t$ is the epoch of the observation (function of observation geometry and illumination conditions). Assuming the above scheme, maps of the rock abundance of lunar terrain were calculated using the discrepancy between nighttime brightness temperatures from Diviner's thermal channels on board of the Lunar Reconnaissance Orbiter (LRO), for fields of view that contain mixtures of warm rocks and cooler regolith. Diviner's short-wavelength thermal infrared detectors measure higher temperatures than the longer-wavelength detectors do for a given scene containing both rocks and fine-grained materials. Through the analysis of three separate wavelengths, \citet{Bandfield2011JGRE..116.0H02B} solved simultaneously for: the areal fraction of each scene occupied by exposed rocks $\sim$1 m and larger; and the temperature of the fines, using modeled rock temperatures as inputs. Theoretically, also the rock temperature could have been estimated given the availability of three channels; practically, the solution of inverse problem is not unique. Because of that, lunar rock temperatures were modeled \textit{a priori} assuming the properties for vesicular basalt described in \citet{1972HoraiSimmons}. The thermal inertia was set to be 1570 \inertiaunit~ at 200 K; the rock had albedo 0.15 and emissivity 0.95 and was modeled as an infinitely thick, level, and laterally continuous layer. The analysis of lunar surface resulted in a very low rock abundance (0.4\% global average within $\pm60^\circ$ latitude) which confirms the dominance of fine regolith due to micrometeorite bombardment and space weathering, expected also from thermal inertia measurements. 

On Mars, infrared remote sensing data have been used for decades to make interpretation of rock abundance \cite[chapter 18 in][and references therein, for a review]{2008Bellbook}, starting in earnest with analyses of the martian surface by infrared radiometers aboard the Mariner probes \citep{1971Neugebauer}, the IRTM (Infrared Thermal Mapper) instruments on the Viking orbiters \citep{1976Kieffer}. Although the physics of heat flow through regolith on Mars differ from airless bodies in that conduction by gas in the pore spaces is quite importance, the same general techniques to interpret infrared emission data may still be applied. As an example, the rock abundance and the physical properties of the regolith fines on Mars have been analyzed on a global scale using IRTM data \citep{1986Christensen,1986JGR_Jakosky} and Thermal Emission Spectrometer (TES) \citep{2000Mellon, 2007Nowicki} by comparing short and long wavelength channels (7 and 20 $\mu m$. and 9 and 30 $\mu m$ for IRTM and TES respectively) for brigthness temperature derived using the KRC 1D model \citep[for a review]{2013Kieffer}. Global thermophysical maps at 100 m/pixel resolution have also been derived from THEMIS (Thermal Emission Imaging System) data \citep{2006Fergason}, although rock abundance has not been estimated due to the limited number of spectral bands (i.e., lack of a high-resolution long-wavelength channel). In-situ, high-resolution thermal analyses have also been conducted on Mars by mini-TES \citep{2006Fergason_b} and REMS \citep{2014HamiltonVic} on board the MER rovers and the MSL rover, respectively. In general, these analyses are in good agreement with their orbital counterpart.

On asteroids, the thermal inertia of regolith and rock components, and relative rock abundance, have never been measured from infrared observations. Infrared data of asteroids are usually acquired from ground-based telescopes. This does not allow to observe the night-side of the asteroid, whose infrared flux is mostly dominated by the emission from rocks. The observed fluxes are matched to surface properties by means of a thermophysical model of the asteroid, which maps input parameters (surface properties, observation geometry, illumination conditions, instrument performances, shape model of the asteroid) into an infrared flux \citep{2015Delbo}. The canonical approach is to generate a large number of cases, i.e., entries of the type \{surface properties, infrared flux\}, and then look for the solution which minimizes the residual error between the predicted and the observed fluxes. Common practice is to look for the minimum of the reduced $\chi^2$, defined as:

\begin{equation}
\label{chi2eq}
    \chi^2_r = \frac{1}{\nu}\sum^{N}_{i=1} \frac{\big(O-M(\textbf{x})\big)^2}{\sigma^2}
\end{equation}
\noindent
where "O" is the observed flux, "$M(\textbf{x})$" is the modeled flux corresponding to the surface properties (array $\textbf{x}$), $\sigma$ is the measurement error on the observed fluxes, N is the number of observations and $\nu$ = N - DOF, where DOF is the number of degrees of freedom (the number of parameters to be estimated). According to the $\chi^2_r$ statistics, all the solutions with $\chi^2_r$ in the range: 

\begin{equation}
\label{range}
  \chi^2_r ~\in ~\big[~\chi^2_{r,min},~ \chi^2_{r,min} + \frac{\sqrt{2\nu}}{\nu}~\big]
\end{equation}
\noindent
are acceptable. If the dimensionality of the problem is enhanced, however, the canonical $\chi^2_r$ approach is unsuccessful. It is indeed computationally expensive to map the parameter space at a resolution sufficient to rule out local minima. To appreciate this, we can think of a model with four inputs (e.g., roughness of the surface, rock abundance, thermal inertia regolith and rock) used to fit infrared fluxes acquired by a nearby spacecraft observing both the day- and night-side of the asteroid. A primary source of computational complexity arises from modelling the surface roughness. This is done by carving craters on the surface for different values of semiaperture angle
and crater surface density; the global effect is an effective surface tilt, also called the macroscopic roughness angle or Hapke angle \citep{1984Hapke}. The calculator needs to compute a full heat exchange solution (i.e., considering all the view factors within the concavities of both the shape model and craters), as the usual \citet{1998Lagerros} approximation cannot be used to fit the night-side observations. With this simulation setting, the asteroid attains simulated thermal equilibrium with the environment in about 3.5 minutes (on a single processor 2.8 GHz Intel Core i7, for a shape model of the asteroids with 2292 facets).  Let us sample the parameter space using 25 points for roughness, 25 for the thermal inertia of regolith and 25 for that of the rock components; the solutions can be linearly combined to study the effect of various rock abundances (Equation \ref{superpos}). The exercise requires to run 1250 full heat exchange thermal simulations (about 3 days). This sampling of the parameter space, however, is too coarse to identify local minima or saddle points of the $\chi^2_r$ function. Therefore, building a look-up table can be misplaced effort, unless the functional relationship between parameters and infrared thermal response is learned and generalized in some way. As we show in the following, the dataset can be instead used to train a neural network capable of predicting the infrared fluxes; as opposed to the ``parent" model, this tool can enable a fine -- and fast -- mapping of the parameter space into the corresponding infrared flux within a known level of accuracy .

\subsection{Surrogate models for thermophysical applications}
\label{ourmethod}

The new approach consists in approximating the thermophysical function $\boldsymbol{y}=\boldsymbol{F}(\boldsymbol{x})$ ($\boldsymbol{x}$: surface properties; $\boldsymbol{y}$: infrared flux) using neural networks trained on a dataset of thermal simulations of the type: \textit{\{surface properties; infrared flux\}}. After training, the networks can predict an infrared flux at at highly reduced computational time with respect to the ``parent" model (running in less than a second versus minutes on a single processor 2.8 GHz Intel Core i7), thus enabling Markov Chain Monte Carlo (MCMC) Bayesian inference of surface properties from observed fluxes. MCMC requires thousands of runs of the forward model to sample the unknown posterior distributions -- thus the need for a fast and accurate predictor, i.e., a trained neural network. 

\begin{enumerate} 
\item A dataset of thermophysical simulations is generated using the TPM code \citep{2015Delbo}, for different combinations of surface properties, given the shape model of the asteroid, the observational geometry and the illumination condition;
\item The dataset is used to train, validate and test a neural network (Section \ref{NN}). The trained network is a surrogate model which predicts the asteroid's infrared flux at a highly reduced computational speed with respect to the "parent" thermophysical model (i.e., less than a second on a single processor 2.8 GHz Intel Core i7).
\item  A series of blind tests is performed to approve the methodology. Each blind test consists in the generation of a case TPM flux by one of the co-author; the case flux is sent to another co-author, who does not know the values of the surface properties used to generate the case. The second co-author performs a Markov Chain Monte Carlo Bayesian inversion of the case TPM flux using the surrogate model (Section \ref{MCMC}). The posterior distributions of the surface properties are sent back to the first co-author -- the only person who knows the values of the surface properties of the case flux -- who verifies the success of the inversion.
\item Once the methodology is validated, we use it to invert observed infrared flux in order to characterize the surface properties of the body, e.g., asteroid (25143) Itokawa in Section \ref{Itokawa}.
\end{enumerate}

\subsubsection{Neural Networks}
\label{NN}
Neural Networks (NN) are mathematical models that take inspiration from the fundamental structure of the brain \citep{schmidhuber2015deep}. NNs are able to learn (i.e., improve the performance of a specific tasks) from data, commonly subdivided in training, validation and testing subsets. NNs are very successful in modeling the functional relationship between inputs and outputs. Such relationship is generally represented by a set of examples (training set) and learned during the training process. NNs can be employed as physically-based fast surrogate models for parameter estimation, replacing computationally expensive numerical models. 

NNs consist of many mathematical units called neurons that are connected via synapses. Neurons are organized by layers and communicate in a parallel fashion through weights that represent the strength of the corresponding synapses. A typical shallow network employed to include surrogate models, comprises one input layer, one hidden layer and an output layer. The hidden layer is assumed to have a specified number of neurons $L$ which are the basic processing units for the network. The overall process begins with a summation of each input with the correspondent weights (synapses) and then further processing by an activation function. In regression problems, the overall NN output function is typically represented as follows:

\begin{equation}
    f_L(\boldsymbol{x}) = \sum_{i=1}^L\boldsymbol{\beta}_i g_i(\boldsymbol{x}) = \sum_{i=1}^L\boldsymbol{\beta}_i G(\boldsymbol{a}_i,b_i,\boldsymbol{x})
\end{equation}
\noindent
where $\boldsymbol{x}\in\ \mathbb{R}^d$ and $\boldsymbol{\beta}_i \in \mathbb{R}^m$. Neurons can usually have different activation functions $g(\cdot)$. Additive nodes with activation functions have the following structure:

\begin{equation}
     G(\boldsymbol{a}_i,b_i,\boldsymbol{x}) = g(\boldsymbol{a}_i^T\boldsymbol{x}+b_i)
\end{equation}
An activation function which works well in shallow neural networks is the tanh-sigmoid function
\begin{equation}
   h(y) = \frac{2}{1+ \exp(-2y)}-1
\end{equation}
\noindent
where $\boldsymbol{a}_i \in \mathbb{R}^m$ and $b_i \in \mathbb{R}$. The weights $\boldsymbol{a}_i$ and biases $b_i$ are usually determined during the training process which implies minimization of a loss function. For regression problems which are usually solved to generate surrogate models, the typical loss function is the Mean Square Error (MSE), i.e.:
\begin{equation}
\label{cost function}
    \epsilon(\boldsymbol{a}_i,b_i) = \frac{1}{N}\sum_{i=1}^N(f_L(\boldsymbol{x}_i)-\boldsymbol{F}(\boldsymbol{x}_i))^2
\end{equation}
where $\{(\boldsymbol{x}_i, \boldsymbol{F}(\boldsymbol{x}_i)\}$ is the associated training set. In our application, $\boldsymbol{x}_i$ is the vector of surface parameters and $\boldsymbol{F}(\boldsymbol{x}_i)$ is the corresponding infrared fluxes. The accuracy of the network is also quantified by the correlation coefficient between predictions and labels. MSE and correlation coefficient are indicator of the reliability with which the network "mimics" (approximate) the "parent" model (e.g., the TPM code).

The dataset is split into a training set (70\%), validation set (15\%) and testing set (15\%). The training set is employed for proper network training, which involves the fitting of the network parameters (weight $\boldsymbol{a}_i$ and biases $b_i$) by minimizing a cost (loss) function (Equation \ref{cost function}). Loss minimization is implemented by standard backpropagation and stochastic gradient descent which are the most common modern approach to training \citep{schmidhuber2015deep}. The overall process is not a simple interpolation of the training set, but rather it involves the search for classes of parametric functions (i.e., the neural network) that globally fit the data (generalization). It is indeed undesirable for any algorithm to match exactly the training set (e.g., through spline interpolation), as the method would then perform poorly on unseen data. The validation set is employed to determine the optimal network hyperparameters (e.g., the learning rate). Finally, the testing set is employed during the training process to evaluate the behavior of the MSE on an unseen ensemble of data, for an independent assessment of the generalization capabilities of the network. Importantly, both validation and testing sets are employed to protect against overfitting of the training set \citep{bishop1995neural}. Indeed, properly trained networks ensure that the data in the validation and the testing sets follow the same probability distribution of the data in the training dataset. Training occurs as a succession of training epochs. At each epoch, the network processes the full set of training data (i.e., it computes the response for all of the data in the training set). Consequently, backpropagation of the residuals between targets and prediction is executed to update the network parameters (weights and biases) that reduce the MSE. Additionally, at every epoch, the MSE for validation and testing is computed; the training is completed (convergence) if the validation MSE does not decrease further after 6 consequent epochs.

\subsubsection{Markov Chain Monte Carlo Bayesian inversion}
\label{MCMC}
The general goal of inverse problems is to use a physical or surrogate model to retrieve properties that characterize a physical system.
In the Bayesian approach, the parameters to be retrieved are treated as random variables. The Bayesian approach to inverse problem generally yields a probability distribution (\textit{a-posteriori} or posterior distribution) for each of the retrieved variables. Such distribution is obtained by virtue of the Bayes' theorem which combines \textit{a-priori} distribution and \textit{likelihood} given the observed data to obtain an estimation of the parameters distribution \citep{stuart2010inverse,aster2011parameter}. Indeed, once observations are collected, one can compute the conditional probability distribution $L(\boldsymbol{m}|\boldsymbol{x})$, i.e. given a certain set of parameters to be retrieved $\boldsymbol{x}$, the set of observations $\boldsymbol{m}$ are observed. As mentioned above, the Bayes' theorem is employed to combine such $likelihood$ conditional probability distribution, $L(\boldsymbol{m}|\boldsymbol{x})$, with the prior distribution $\pi_{pr}(\boldsymbol{x})$ to compute the conditional posterior distribution for the model parameters $\pi(\boldsymbol{x}|\boldsymbol{m})$. Solving inverse problems within the Bayesian framework relies on computing the conditional posterior probability distribution, i.e.:

\begin{equation}
    \pi(\boldsymbol{x}|\boldsymbol{m}) = \frac{\pi_{pr}(\boldsymbol{x})L(\boldsymbol{m}|\boldsymbol{x})}{\pi(\boldsymbol{m})}
\end{equation}
\noindent
where $\pi(\boldsymbol{m})$ is the probability distribution of the observed data:
\begin{equation}
	\pi(\boldsymbol{m}) =\int_x \pi(\boldsymbol{x},\boldsymbol{m})~dx = \int_x \pi_{pr}(x) L(\boldsymbol{m}|\boldsymbol{x})~dx
\end{equation}
\noindent
Generally, the inverse problem is regularized by using the prior distribution $\pi_{pr}(\boldsymbol{x})$ which encodes prior knowledge or belief about the parameters. As stated in \cite{stuart2010inverse}, since all variables are assumed to be random, they reflects the uncertainty of the true values and, importantly, the degree of uncertainty is encoded in the probability distribution of the parameters to be retrieved.
Typical estimates using $\pi(\boldsymbol{x}|\boldsymbol{m})$ are the 1) Maximum A-Posteriori (MAP) distribution and 2) Mean of the posterior distribution. If the posterior distribution is normal, the MAP and posterior mean will coincide \citep{stuart2010inverse,aster2011parameter}. Generally, analytical evaluations of such integrals are quite impossible. When $n$ is large, traditional numerical quadrature schemes are not practical and one has to resort to Monte Carlo (MC) integration methods. MC algorithms are applied by sampling $N$ points on $\pi(\boldsymbol{x}|\boldsymbol{m})$, ${\boldsymbol{x}_k, k = 1,2,...,N}$. In this work, a Monte Carlo Markov Chain (MCMC) approach is employed for sampling. There are many algorithms for MCMC; one of the most widely used is the Metropolis--Hastings (MH) algorithm that performs a Markov chain with a specified limiting distribution. Other used algorithms for MCMC are the Adaptive Metropolis-Hastings (AM), Delayed rejection (DR), or their combination called DRAM \citep{haario2006dram}.\\

We demonstrate the potentialities of the proposed method by applying it to two case studies. The first application (Section \ref{Bennu}) discusses the inversion of simulated infrared fluxes of asteroid (101955) Bennu as acquired by the OSIRIS-REx mission. The second application (Section \ref{Itokawa}) is about the inversion of observed infrared fluxes of asteroid (25143) Itokawa. In both cases, the method is able to accomplish the inversion and to provide insights about the surface properties of the bodies.

\section{Results and Discussion}
\subsection{Blind tests on simulated infrared flux of (101955) Bennu}
\label{Bennu}
Once at the target asteroid (101955) Bennu, the OSIRIS-REx mission will perform a detailed survey of the surface in order to constrain its properties and select the sampling site \citep{2017LaurettaOrex}. OSIRIS-REx requirement on sampling is to collect regolith not hotter than  350K, and TAGSAM should be used to collect fine regolith, e.g., grain size less than 2 cm. Furthermore, OSIRIS-REx should avoid rocks (which are hazardous) during the sampling phase. These mission requirements will constrain  the  choice  of  the latitude of the sample selection area, the local time, and the arrival date at the asteroid \citep{2015Delbo}.  The OSIRIS-REx Thermal Emission Spectrometer \citep[OTES,][]{2018ChristensenOTES} will collect the infrared fluxes emitted by the surface of the asteroid covering the spectral range  $5.71-100~\mu m$ $(1750-100~cm^{-1})$ with a spectral sample interval of $8.66~cm^{-1}$. The thermal properties of the surface will inform the team about the sampleability of the different regions of the asteroid. The abundance of rock and surface roughness will be used by the mission engineers to define the hazard associated with sampling a certain region; the thermal inertia of the regolith can be converted in average grain size \citep{2013Gundlach,Sakatani2017AIPA....7a5310S}.

\subsubsection{Surrogate models for OTES acquisition} 
\label{OTES_surrogate}

In the case of Bennu, we apply the methodology described in Section \ref{method} as follows. A 4-dimensional dataset of infrared fluxes (4 surface properties: surface roughness, thermal inertia of the regolith, thermal inertia of the rock, rock abundance) is populated with simulated OTES-like acquisitions of the infrared flux from (101955) Bennu, using the TPM \citep{2015Delbo}. We work with two shape models: a single triangular equatorial facet, and the full shape model of the asteroid by \citet{2013Nolanshape}. The shape model is represented by a mesh of 2292 triangular facets. Facets can have any number of vertices, in principle; however, the temperature and flux of the facet depends only on its area and normal direction. On the surface, we use the full heat exchange solution (i.e., considering all the view factors within the concavities of both the shape model and craters), as the \citet{1998Lagerros} approximation is not suitable to fit the night-side observations. In the simulations, the observer is always looking nadir, i.e. the unity vector from the center of the asteroid to the observer and the normal to the surface have scalar product equal to one, and it is located at infinite distance on the equatorial plane. The heliocentric distance is assumed to be 1 au. The model output are 7 fluxes in the wavelength range between 5 and 50 $\mu m$, each corresponding to one of the different O-REx's equatorial stations of the detailed survey (03:20 AM, 06:00 AM, 10:00 AM, 12:30 PM, 03:00 PM, 06:00 PM, 08:40 PM). We model the surface with surface roughness values between 4 and 62 degrees \citep[in terms of Hapke angle,][]{1984Hapke}. We assume constant thermal inertia on the surface, for values varying between 0 and 1000 \inertiaunit. The fluxes corresponding to values of thermal inertia below 500 \inertiaunit~are assumed to be emitted by a surface covered in regolith ($f_{regolith}$). The fluxes corresponding to values of thermal inertia above 500 \inertiaunit~are instead assumed to be emitted by a surface covered in boulders ($f_{rock}$). The final dataset is composed by all the possible combinations of $f_{regolith}$ and $f_{rock}$, weighted in terms of rock abundance (see Equation \ref{superpos}, Table \ref{Bennudat}). Building the dataset requires to run 1200 TPM simulations (12 different values of roughness for 100 different values of thermal inertia). Each simulation runs in 0.602 seconds for a single facet and 3.5 minutes for the disk-integrated case, corresponding to about 12 minutes and 3 days for the overall building of the dataset, respectively. The fluxes for regolith and rock are then linearly combined in terms of rock abundance (Equation \ref{superpos}). 

The dataset of infrared fluxes is used to train seven (7) surrogate models (one for each of the equatorial station) which mimic the acquisition of infrared fluxes at Bennu by OTES. Each of the surrogate model is derived from the deployment of shallow neural networks on the 4-dimensional dataset of infrared simulated fluxes (Table \ref{Bennudat}). We refer to Section \ref{NN} for the training, validation and testing procedure, and to Table \ref{bennuperf} for the architecture and performance of the networks. The whole training procedure requires a computational effort of about 10 minutes. The trained networks run in less than a second. All the run time are computed on a single processor 2.8 GHz Intel Core i7. 

\begin{figure}[h!]
\centering
  \includegraphics[width=\linewidth]{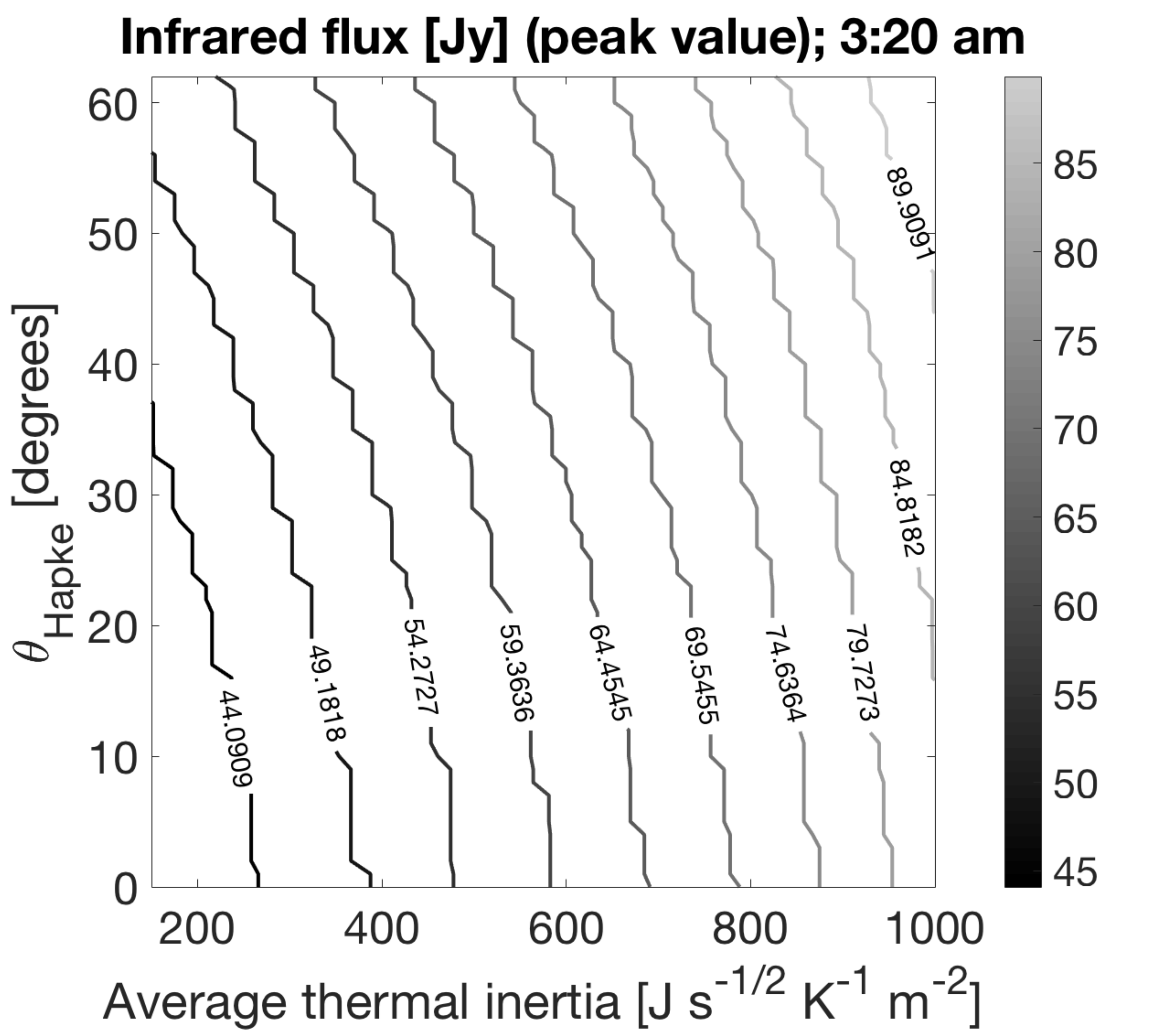}
  \caption{Iso-flux curves (peak value of the infrared flux, in Jy), of (111955) Bennu as recorded from the 3:20 am equatorial station during the detailed survey. The fluxes are predicted using the corresponding neural network (see the text in Section \ref{OTES_surrogate} for details). Higher fluxes are due to higher values of average thermal inertia (e.g., higher rock abundance in the surface) and/or enhanced surface roughness \citep[self-heating by re-absorption of thermal radiation in craters,][]{2004Delbophd}.}
  \label{var_par}
\end{figure}

Each neural network maps the flux into a 4-Dimensional parameter space (surface roughness, thermal inertia of the regolith, thermal inertia of the rock, rock abundance). As an example, Figure \ref{var_par} is a contour plot of the peak value of the infrared flux, as predicted by the neural network and observed on the night-side (equatorial station: 3:20 am). The free parameters are surface roughness and the average thermal inertia in lieu of rock abundance (RA), e.g.,

\begin{equation}
\label{gamma_tot}
    \Gamma_{average}=\Gamma_{rock}~RA+\Gamma_{regolith}~(1-RA)
\end{equation}
\noindent 
where $\Gamma_{rock}$ and $\Gamma_{regolith}$ are the thermal inertias of the rock and the regolith. In the example of Figure \ref{var_par}, $\Gamma_{rock}$ and $\Gamma_{regolith}$ are kept constant -- and equal to 1000 and 150 \inertiaunit, respectively -- to allow a 2-D visualization of the parameter space. The peak flux increases with the average thermal inertia because rocks remain warmer on the night-side and, the higher the rock abundance, the higher the average thermal inertia, the higher the peak flux. An increase in the flux is also observed with roughness. Since roughness is modeled by carving craters on the surface, we observe an increase in flux because of self-heating by re-absorption of thermal radiation from other parts of the crater \citep[Figure 6.8 in][]{2004Delbophd}.

\begin{table*}[h!]
\centering
\begin{tabular}{ |c|c|  }
 \hline
 \multicolumn{2}{|c|}{\textbf{Dataset of forward simulations (OTES-like acquisitions)}} \\
 \hline
 \textbf{Predictors} & \textbf{Values} \\
 \hline
 Hapke angle   & 4, 12, 14, 21, 27, 30, 34, 40,  46, 50, 56, 62 degrees   \\
 Thermal inertia, reg.  &   10 : 10 : 500 \inertiaunit \\
 Thermal inertia, rock  &  500 : 10 : 1000  \inertiaunit \\
 Rock abundance     & 0 : 5 : 100 \% \\
\hline
 \textbf{Responses} & \textbf{Values} \\
 \hline
 7 infrared fluxes   & [$Jy$], 5  to  50 $\mu m$ with  a  step  of  1 $\mu m$   \\
 \hline
\end{tabular}
\caption{Description of the dataset of forward thermal simulations of asteroid (101955) Bennu. The values of thermal inertia and rock abundance are equally spaced in a range; as an example, the diciture ``10 : 10 : 500 \inertiaunit~" means ``values from 10 to 500 \inertiaunit~with a step of 10 \inertiaunit".  Each thermal simulation uses a spin state: $\lambda$ = $0^\circ$, $\beta$ = 90$^\circ$, $P$ = 4.288 hours. The asteroid is located at 1 au.  The asteroid is thermalized, i.e., it attains thermal equilibrium with the environment, in over several tens of rotational periods before recording the infrared flux. Namely, we begin each simulation where each facet has a surface temperature equal to its mean temperature calculated over the full diurnal curve and assuming zero thermal inertia. The temperature in the sub-surface of the facet is initially the same of the surface temperature. Next the appropriate value of the thermal inertia for the simulation is set and the heat diffusion into depth is calculated using Equation 12 of \cite{Spencer1989Icar...78..337S} with the boundary conditions given as in Equation 13 of \cite{Spencer1989Icar...78..337S}. The simulations are started at an epoch before the one when infrared fluxes are needed to be modeled and the temperatures of the facet (or some facets) are monitored such that the diurnal temperature curve at the asteroid rotation $N$ is within a fraction (we use typically 0.25 K) of a K from the diurnal temperature at the rotation $N+1$. The number of "thermalization" rotations is determined by trial and error.}
\label{Bennudat}
\end{table*}

\subsubsection{MCMC Bayesian inversion}

The reduced computational time of the neural networks and their capability of generalizing a response for predictors in between the training nodes make these models statistically invertible (Section \ref{MCMC}). The methodology is here blind-tested against a set of case fluxes (step 3 in Section \ref{ourmethod}). The case flux is generated using the TPM code and mimic a "real" observed flux that could be acquired by OTES.  The fluxes' Signal-to-Noise Ratio (SNR) is modeled such that it resembles the real performance of the spectrometer \citep[and priv. comm.]{2018ChristensenOTES}. We use the trained networks as forward models in a Markow Chain Monte Carlo Bayesian inversion of the simulated infrared fluxes. For this exercise, the inversion is not informed about any \textit{a-priori} information (i.e., we use uninformative \textit{a-priori} distributions). In real application, it is reasonable to have some \textit{a-priori} information on the parameters, e.g., the OSIRIS-REx Visible and Infrared Spectrometer \citep[OVIRS,][]{2018OVIRS} would also provide information about the material on the surface, thus providing information about the thermal inertia. Using proper \textit{a-priori} distributions improves the accuracy of the inversion and helps achieving convergence. As we show below, however, for this application the method is able to correctly perform the inversion also without any \textit{a-priori} information.

\begin{table}[h!]
\centering
\begin{tabular}{
|c|c|c|c|  }
 \hline
 \multicolumn{4}{|c|}{\textbf{Surrogate Models: Triangular Equatorial Facet}} \\
 \hline
 Station n. & Hidden layers & MSE & R\\
 \hline
3:20 AM   & 1 layer, 20 neurons & 0.53 &   0.999\\
6:00 AM   & 1 layer, 20 neurons & 0.42   & 0.999\\
10:00 AM   & 1 layer, 10 neurons & 0.90 &  0.999\\
12:30 PM   & 1 layer, 10 neurons & 2.00 &  0.999\\
3:00 PM   & 1 layer, 10 neurons &  0.90 & 0.999\\
6:00 PM   & 1 layer, 10 neurons & 0.30   & 0.999\\
8:40 PM   & 1 layer, 10 neurons & 0.90   & 0.999\\
 \hline
 \hline
 \multicolumn{4}{|c|}{\textbf{Surrogate Models: Disk-integrated}} \\
 \hline
 Station n. & Hidden layers & MSE & R\\
 \hline
 03:20 AM   & 1 layer, 10 neurons & 0.085 &   0.999\\
 06:00 AM   & 1 layer, 10 neurons  & 0.037   & 0.999\\
 10:00 AM   & 1 layer, 10 neurons & 0.053 &  0.999\\
 12:30 PM   & 1 layer, 10 neurons & 0.124 &  0.999\\
 03:00 PM   & 1 layer, 10 neurons  &  0.182 & 0.999\\
 06:00 PM   & 1 layer, 10 neurons  & 0.066   & 0.999\\
 8:40 PM   & 1 layer, 10 neurons  & 0.082   & 0.999\\
 \hline
\end{tabular}
\caption{Architecture and performance of the surrogate models (neural networks) for OTES observations. The performance of the neural networks are assessed in terms of Mean Square Error (MSE, the average squared difference between outputs and targets; lower values are better) and Regression R value (the correlation between outputs and targets; an R value of 1 means a close relationship). The reported values are relative to the testing phase. We used the Matlab Levenberg-Marquardt algorithm for training, with 10-fold crossvalidation. The dataset is split as: 70\% training, 15\% validation and 15\% testing.}
\label{bennuperf}
\end{table}

\begin{figure}[h!]
\centering
  \includegraphics[width=\linewidth]{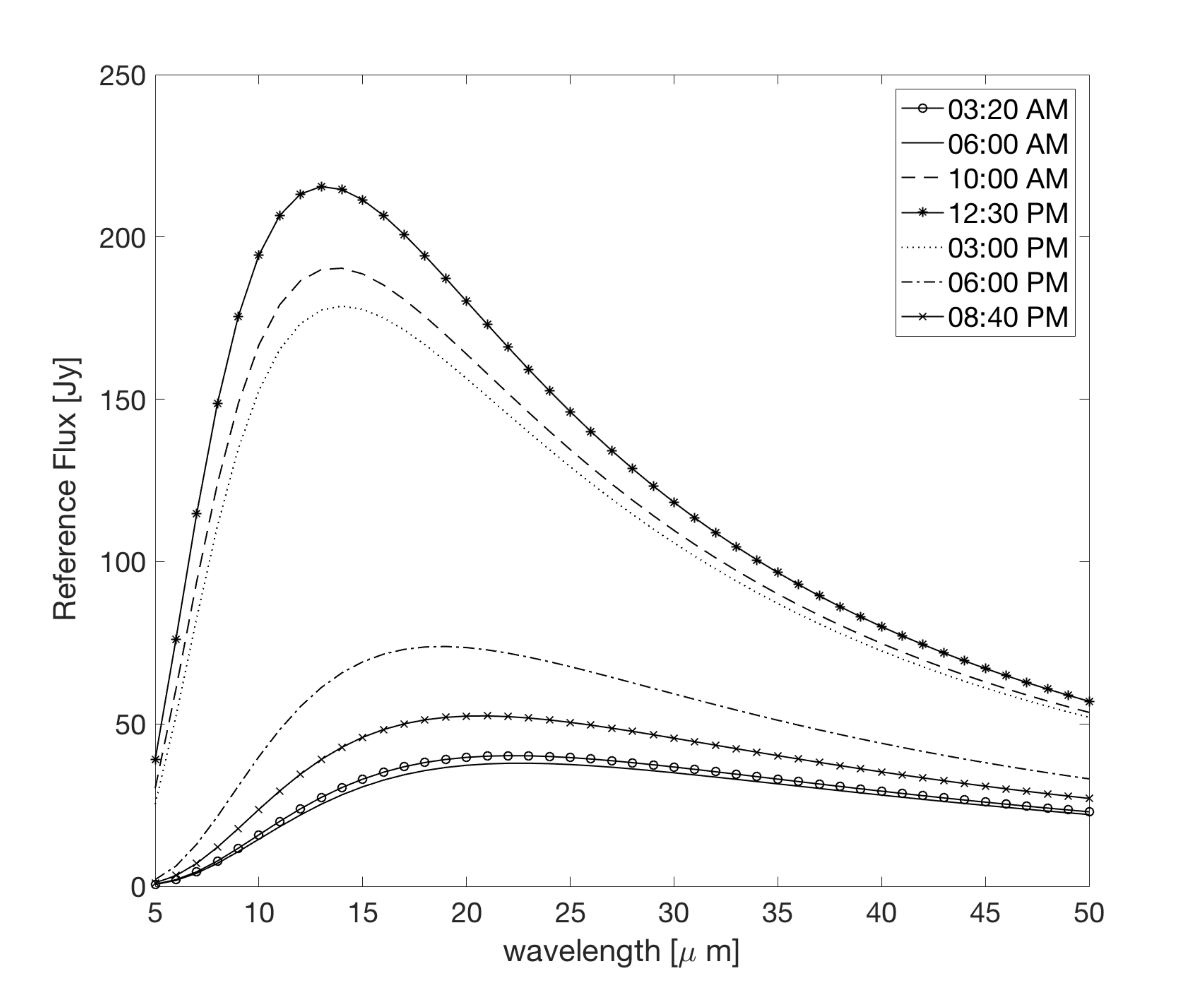}
  \includegraphics[width=\linewidth]{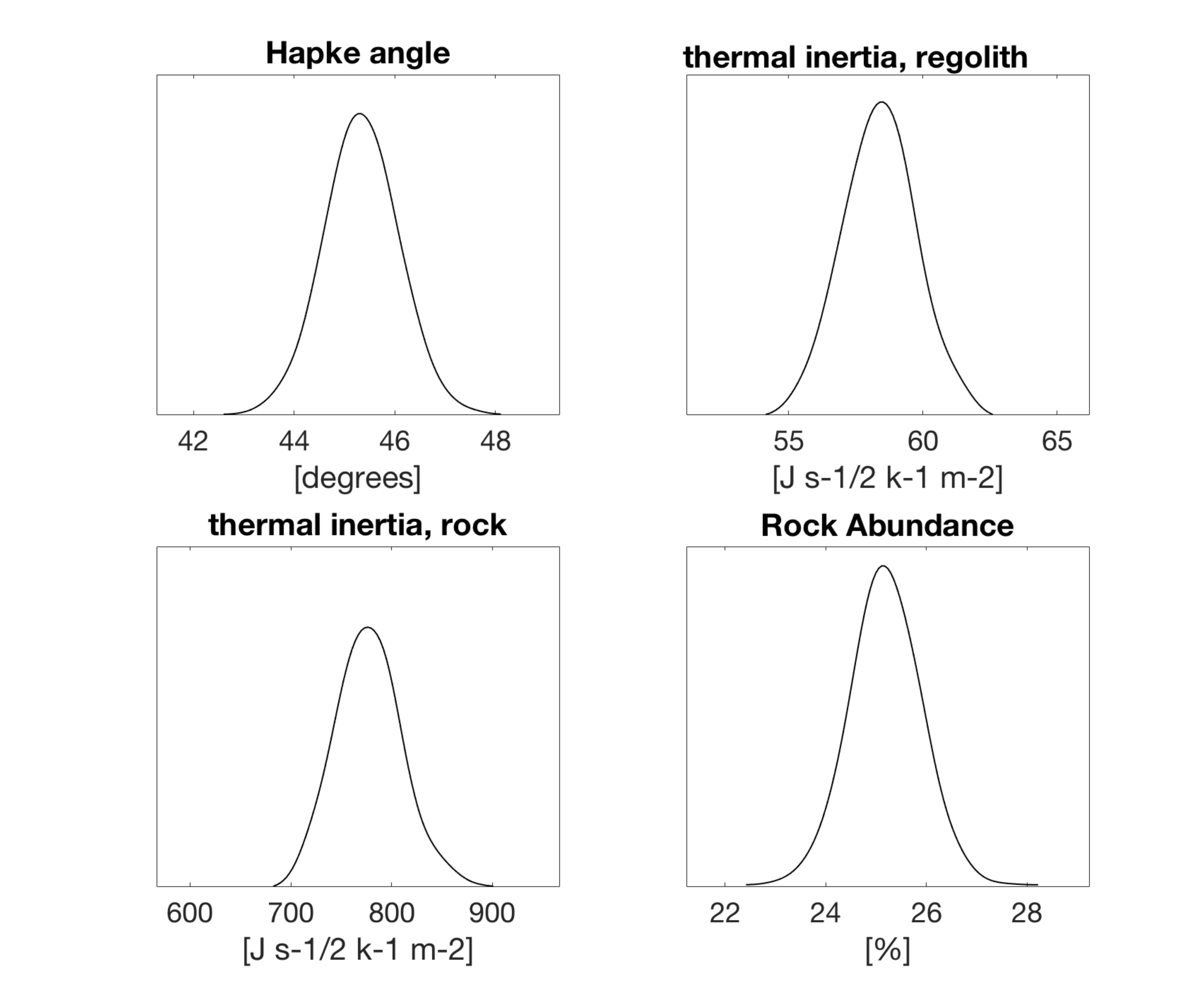}
  \caption{\textit{Triangular equatorial facet}. Top panel: set of reference fluxes acquired at the 7 equatorial stations, for values of the surface parameters: $\bar{\theta} = 46^\circ$,  $\Gamma_{regolith}=60$ \inertiaunit, ~ $\Gamma_{rock}=820$ \inertiaunit, ~ $RA~=~25.0\%$. Bottom panel: posterior distributions of the surface thermal properties; the retrieved surface properties are: $\bar{\theta} = 45.3~\pm~0.7^\circ$,  $\Gamma_{regolith}=58~\pm~2$ \inertiaunit, $\Gamma_{rock}=780~\pm~39$ \inertiaunit, RA = $25.2~\pm~0.7\%$. The reference values for the parameters belong to the posterior distributions.}
  \label{ref_tri}
\end{figure}

\begin{figure}[h!]
\centering
  \includegraphics[width=\linewidth]{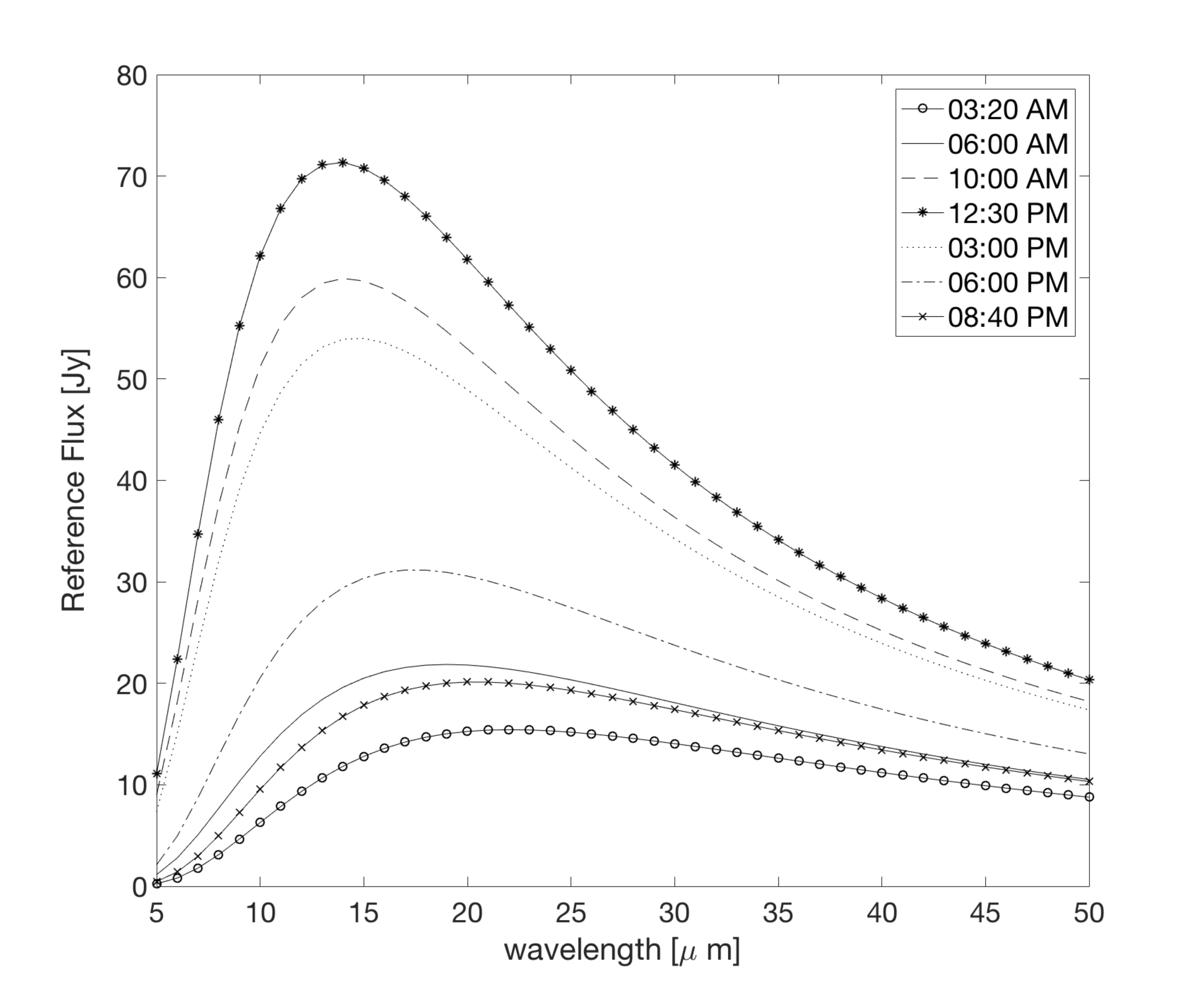}
  \includegraphics[width=\linewidth]{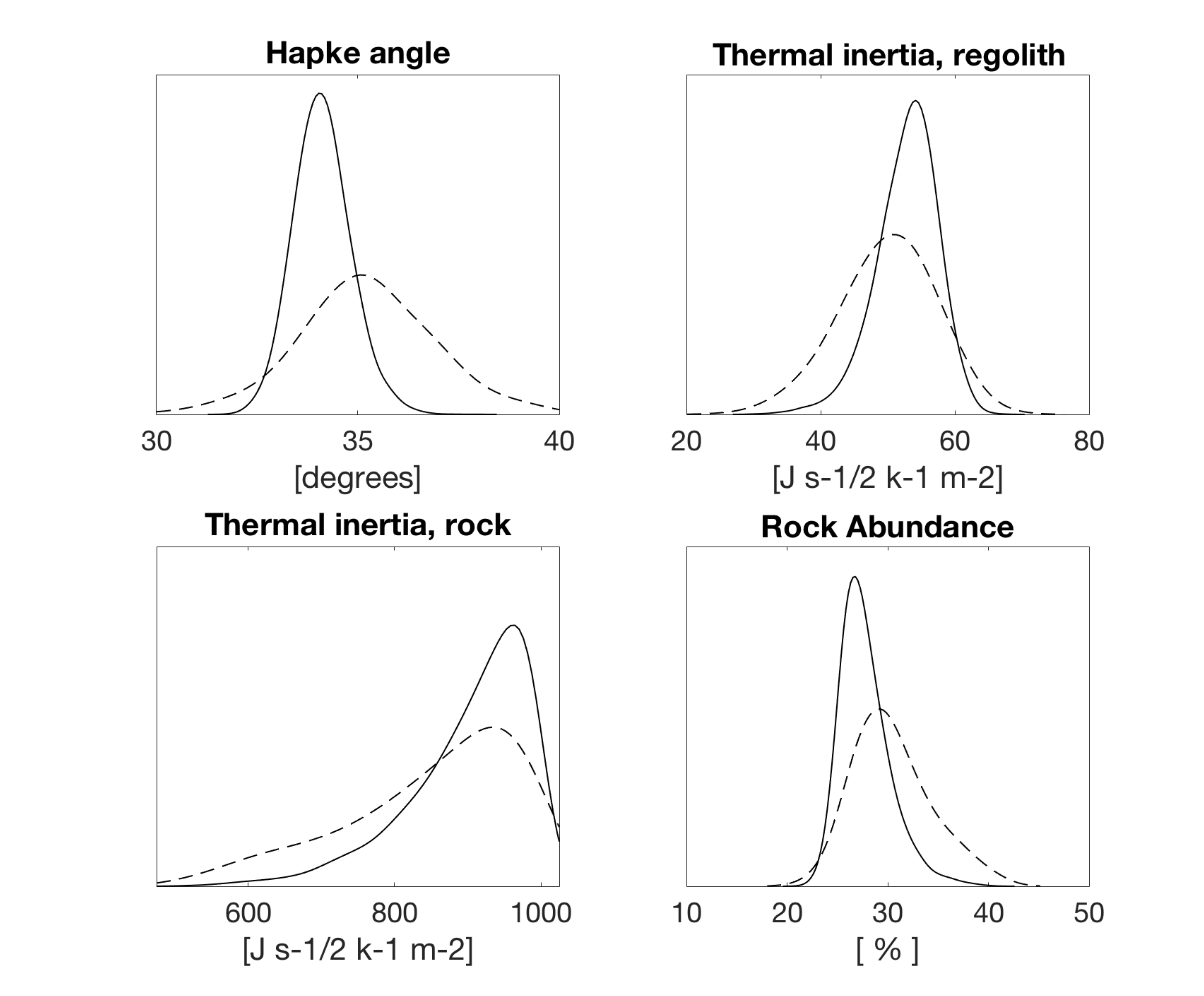}
  \caption{\textit{Disk-integrated fluxes, (101955) Bennu}. Top panel: set of reference fluxes acquired at the 7 equatorial stations, for nominal values of the surface parameters: $\bar{\theta} = 34^\circ$,  $\Gamma_{regolith}=50$ \inertiaunit, $\Gamma_{rock}=850$ \inertiaunit, RA = $30.0\%$.  Bottom panel: posterior distributions of the surface thermal properties. The solid curves are the solutions when all the observations are processed. The retrieved surface properties are: $\bar{\theta} = 34.1~\pm~0.7^\circ,~\Gamma_{regolith}=53~\pm~4$ \inertiaunit,~$\Gamma_{rock}=905~\pm~84$ \inertiaunit, RA = $27~\pm~3\%$. The dashed curves are the posterior distributions of the surface thermal properties when only the daytime observations are processed. In both cases, the reference values for the parameters belong to the posterior distributions; the surface properties are retrieved without using any \textit{a-priori} information.}
  \label{ref_bennu}
\end{figure}

The blind test for the triangular equatorial facet uses a case TPM flux corresponding to surface properties: $\bar{\theta} = 46^\circ$,  $\Gamma_{regolith}=60$ \inertiaunit, $\Gamma_{rock}=820$ \inertiaunit, RA = $25.0\%$ (top panel in Figure \ref{ref_tri}). The MCMC inversion of the blind test provides the posterior distributions in Figure \ref{ref_tri}, bottom panel. The same blind-test procedure, but for the disk-integrated flux, used a case flux corresponding to surface properties: $\bar{\theta} = 34^\circ$,  $\Gamma_{regolith}=50$ \inertiaunit, $\Gamma_{rock}=850$ \inertiaunit, RA = $30.0\%$ (top panel in Figure \ref{ref_bennu}); the posterior distributions from the inversion are in Figure \ref{ref_bennu}, bottom panel. In both cases, the blind tests are successful as the values of the surface properties for the case fluxes belong to the posterior distributions; the difference between the reference and the statistical mean is less than 1--$\sigma$. An exception is the thermal inertia of the rock abundance for the single triangular facet, for which the reference value is within 2--$\sigma$ of the estimated mean value. 

To test the robustness of the method, we additionally perform an inversion using the observations relative to the daytime stations only (10:00 AM, 12:30 AM, 3:00 AM; disk integrated case). On the night-side, the infrared flux is mainly defined by the rocky material because of its inherently higher thermal inertia and no shadow is present. Therefore the nighttime observations are important to constrain the presence of rocky material and to get a more complete characterization of the surface. The posterior distributions for this case are the dashed curves in Figure \ref{ref_bennu}, bottom panel. The inversion of the disk-integrated case using only the daytime stations is still successful. The main difference with respect to the full inversion is observed on the posterior distribution of the thermal inertia of the rock; this result was expected as the rocks appear brighter in the infrared on the night-side (as it cools down slower than the regolith). We conclude that the inversion method is able to retrieve the correct surface properties both for a single facet and for the disk-integrated case without using any \textit{a-priori} information.  The (small) deviations between the reference and retrieved values are attributed to the effect of noise from the OTES instrument. 

\subsection{Inversion of observed infrared fluxes of (25143) Itokawa}
    \label{Itokawa}
    
The near-Earth asteroid (25143) Itokawa was characterized in great detail by both ground-based observations and the JAXA's
Hayabusa mission. Here we use the proposed methodology to invert the available mid-infrared photometric observations and we compare the results with pre and post-Hayabusa findings, e.g., surface roughness, thermal inertia and grain size of the regolith \citep{2008kitazato, 2014MuellerT, Sakatani2017AIPA....7a5310S}. The mid-infrared observations are summarized in Table \ref{obs_ito}. These consist in 25 ground-based observations and five measurements from the Japanese
infrared astronomical satellite AKARI  \citep{2005MuellerT,2014MuellerT}. All the epoch of observations are the time of emission by the asteroid. The position vectors of the asteroid and the Earth have been queried to the JPL Horizons Ephemeridis system (\url{https://ssd.jpl.nasa.gov/horizons.cgi}) and converted in the asteroid-centric reference frame. 

\begin{table}[h]
\centering
\begin{tabular}{
|c|c|c|c|c|  }
 \hline
 \multicolumn{5}{|c|}{\textbf{Thermal observations of (25143) Itokawa}} \\
 \hline
 N. & Epoch [JD] & $\lambda$ [$\mu m$] & Flux [$Jy$] & $\sigma$ [$Jy$]\\
 \hline
  1	&	2451982.743	&	11.66	&	0.264	&	0.044	\\
2	&	2452007.894	&	11.66	&	0.164	&	0.021	\\
3	&	2452007.904	&	10.38	&	0.144	&	0.018	\\
4	&	2452007.917	&	12.35	&	0.17	&	0.022	\\
5	&	2452007.929	&	8.73	&	0.086	&	0.022	\\
6	&	2452007.94	&	11.66	&	0.149	&	0.019	\\
7	&	2452008.894	&	12.35	&	0.258	&	0.032	\\
8	&	2452008.906	&	9.68	&	0.108	&	0.016	\\
9	&	2452008.919	&	10.38	&	0.169	&	0.027	\\
10	&	2452008.929	&	11.66	&	0.242	&	0.03	\\
11	&	2452008.939	&	11.66	&	0.193	&	0.028	\\
12	&	2453187.752	&	8.73	&	1.92	&	0.15	\\
13	&	2453187.763	&	8.73	&	1.97	&	0.16	\\
14	&	2453187.775	&	8.73	&	1.75	&	0.14	\\
15	&	2453187.788	&	8.73	&	1.67	&	0.13	\\
16	&	2453187.803	&	10.68	&	1.94	&	0.14	\\
17	&	2453187.817	&	10.68	&	1.89	&	0.13	\\
18	&	2453187.828	&	12.35	&	2.17	&	0.13	\\
19	&	2453187.84	&	12.35	&	1.8	&	0.11	\\
20	&	2453187.859	&	17.72	&	2.49	&	0.5	\\
21	&	2453196.989	&	11.7	&	0.762 	&	0.100 	\\
22	&	2453196.991	&	11.7 	&	0.721 	&	0.091 	\\
23	&	2453197.064	&	11.7 	&	0.913 	&	0.114 	\\
24	&	2453197.07	&	9.8 	&	0.791 	&	0.125 	\\
25	&	2453197.077	&	9.8 	&	0.570 	&	0.122 	\\
26	&	2454307.977	&	4.1 	&	0.00032 	&	0.00025 	\\
27	&	2454307.977	&	7.0 	&	0.00469 	&	0.00028 	\\
28	&	2454307.976	&	11.0 	&	0.01422 	&	0.00053 	\\
29	&	2454308.046	&	15.0 	&	0.02137 	&	0.00079 	\\
30	&	2454308.048	&	24.0 	&	0.01947 	&	0.00120 	\\

 \hline
\end{tabular}
\caption{Mid-infrared observations of (25143) Itokawa from \citet{2005MuellerT, 2014MuellerT}. The epoch of observations are relative to the emission from the surface of the asteroid (i.e., light-time corrected). }
\label{obs_ito}
\end{table}
      
\subsubsection{Thermal surrogate model of (25143) Itokawa} 

The steps for the design of the surrogate model in the case of (25143) Itokawa are analogous to those for (111995) Bennu (Section \ref{OTES_surrogate}), with some proper adaptations which are described in the followings. For (25143) Itokawa we use the TPM code to populate two datasets: a 2-D dataset and a 4-D dataset, Table \ref{itodat}. The infrared fluxes in the 2-D dataset are modeled with two surface properties: Hapke angle and average thermal inertia of the surface. The infrared fluxes in the 4-D dataset are modeled with four surface properties: Hapke angle, thermal inertia of the regolith, thermal inertia of the rock, rock abundance. For the 4-D dataset, the thermal infrared radiation is computed as superposition of the contribution from the regolith and the rock (Equation \ref{superpos}). Each of the infrared fluxes is generated using the shape model by \citep{2008Gaskell} at a resolution of 4096 facets. The surface roughness is represented using craters and the fluxes have been computed using the Lagerros' approximation \citep{1998Lagerros}, as no night-side observations are available. Each run simulates the acquisition of the flux corresponding to the wavelength and epoch of the observation in Table \ref{obs_ito}; the asteroid is observed from Earth. The run time for a simulation is 15.2 seconds if the \citet{1998Lagerros} approximation is employed. 

Each surrogate model is tailored to the specific observation geometry and spectral range and derives from the deployment of shallow neural networks on the two datasets described in Table \ref{itodat}. For each observation in Table \ref{obs_ito}, a neural network is trained, validated and tested on the 2-D dataset, composed by 720 predictions of infrared fluxes corresponding to different combinations of surface properties. Therefore, the total computation cost to populate the 2-D dataset is about 4 days ($720\times30\times15.2$ seconds). The same procedure is followed for the 4-D dataset, which is composed by 1200 simulations per observation (total run time of 5 days). We refer to Section \ref{NN} for the training, validation and testing procedure. The whole training procedure requires a computational effort of about 1 hour. Table \ref{itoperf} indicates the metrics of performance of the networks, which run in less than a second. All the time are computed on a single processor 2.8 GHz Intel Core i7.

\begin{table*}[h!]
\centering
\begin{tabular}{|c|c|}
\hline
 \multicolumn{2}{|c|}{\textbf{2-D model (Itokawa)}} \\
 \hline
 \textbf{Predictors} & \textbf{Values} \\
 \hline
 Hapke angle   & 4, 12, 14, 21, 27, 30, 34, 40,  46, 50, 56, 62 degrees   \\
 Thermal inertia, av.  &   25: 25 : 1500 \inertiaunit\\
\hline
 \textbf{Responses} & \textbf{Values} \\
 \hline
 Infrared flux   & 30x1 fluxes $[Jy]$  \\
 \hline
 \end{tabular}

 \begin{tabular}{ |p{3cm}|p{5cm}|  }
 \hline
 \multicolumn{2}{|c|}{\textbf{4-D model (Itokawa)}} \\
 \hline
 \textbf{Predictors} & \textbf{Values} \\
 \hline
 Hapke angle   & 4, 12, 14, 21, 27, 30, 34, 40,  46, 50, 56, 62 degrees   \\
 Thermal inertia, reg.  &   25 : 25 : 700 \inertiaunit\\
 Thermal inertia, rock  &  700 : 25 : 2500 \inertiaunit \\
 Rock abundance     & 0 : 5 : 100 \% \\
\hline
 \textbf{Responses} & \textbf{Values} \\
 \hline
 Infrared flux   & 30x1 fluxes $[Jy]$  \\
 \hline
\end{tabular}
\caption{Description of the datasets of forward simulations for the thermal model of asteroid (25143) Itokawa. The prime meridian is defined according to the IAU specifications \citep{2008Gaskell}. Each thermal simulation uses a spin state: $\lambda$ = 268.77$^\circ$, $\beta$ = -89.53$^\circ$, $P$ = 12.1324 hours. The asteroid is thermalized in over thirty rotational periods before recording the infrared flux.}
\label{itodat}
\end{table*}

\begin{table}[h!]
\centering
\begin{tabular}{
|p{0.5cm}|p{1.5cm}|p{1.25cm}|p{1.5cm}|p{1.25cm}| }
 \hline
 \multicolumn{5}{|c|}{\textbf{Surrogate models for (25143) Itokawa}} \\
\hline
 \textbf{N.} & \textbf{MSE (2D)} & \textbf{R (2D)} & \textbf{MSE (4D)} & \textbf{R (4D)}\\
 \hline
1	&	1.0E-07	&	0.999	&	1.5E-05	&	0.998	\\
2	&	3.0E-07	&	0.999	&	3.0E-05	&	0.998	\\
3	&	6.0E-08	&	0.999	&	1.5E-05	&	0.998	\\
4	&	5.0E-08	&	0.999	&	1.5E-05	&	0.998	\\
5	&	3.8E-08	&	0.999	&	5.4E-06	&	0.991	\\
6	&	5.0E-08	&	0.999	&	9.0E-06	&	0.991	\\
7	&	1.4E-07	&	0.999	&	3.9E-05	&	0.991	\\
8	&	5.0E-07	&	0.999	&	1.2E-05	&	0.991	\\
9	&	1.3E-07	&	0.999	&	1.5E-05	&	0.991	\\
10	&	1.3E-07	&	0.999	&	1.5E-05	&	0.991	\\
11	&	2.6E-07	&	0.999	&	1.0E-05	&	0.999	\\
12	&	7.8E-08	&	0.999	&	4.0E-04	&	0.999	\\
13	&	4.0E-07	&	0.999	&	6.0E-04	&	0.999	\\
14	&	6.0E-07	&	0.999	&	3.0E-04	&	0.999	\\
15	&	3.0E-07	&	0.999	&	1.0E-04	&	0.999	\\
16	&	2.9E-07	&	0.999	&	1.0E-04	&	0.999	\\
17	&	3.0E-07	&	0.999	&	4.0E-04	&	0.999	\\
18	&	2.0E-07	&	0.999	&	5.0E-04	&	0.999	\\
19	&	9.7E-08	&	0.999	&	1.0E-04	&	0.999	\\
20	&	2.0E-07	&	0.999	&	5.0E-05	&	0.999	\\
21	&	4.0E-07	&	0.999	&	9.8E-06	&	0.999	\\
22	&	8.0E-08	&	0.999	&	8.5E-06	&	0.999	\\
23	&	3.8E-08	&	0.999	&	5.0E-05	&	0.999	\\
24	&	1.2E-07	&	0.999	&	4.0E-05	&	0.999	\\
25	&	4.0E-08	&	0.999	&	1.0E-05	&	0.999	\\
26	&	3.9E-09	&	0.999	&	7.0E-08	&	0.999	\\
27	&	8.2E-09	&	0.999	&	3.0E-08	&	0.999	\\
28	&	3.0E-09	&	0.999	&	1.0E-08	&	0.991	\\
29	&	1.4E-08	&	0.999	&	3.0E-07	&	0.910	\\
30	&	1.3E-08	&	0.999	&	2.0E-07	&	0.900	\\
\hline
\end{tabular}
\caption{Metrics of performance of the surrogate models (neural networks) for the thermal observations of (25143) Itokawa, deployed on the 2-D dataset and on the 4-D dataset. For the 2-D case, all the networks have a single hidden layer with 30 neurons. For the 4-D case, all the networks have a single hidden layer with 10 neurons, except networks n. 17 and 20, that have 20 neurons. The reported values are relative to the testing phase.  The average Mean Squared Error of the surrogate models is within the noise level of the observations. We used the Matlab Levenberg-Marquardt algorithm for training, with 10-fold crossvalidation. The dataset is used as: 70\% training, 15\% validation and 15\% testing.}
\label{itoperf}
\end{table}

\subsubsection{Bayesian inversion of the 2-D dataset}
\label{ito2D}

\begin{figure}[h!]
\centering
  \includegraphics[width=\linewidth]{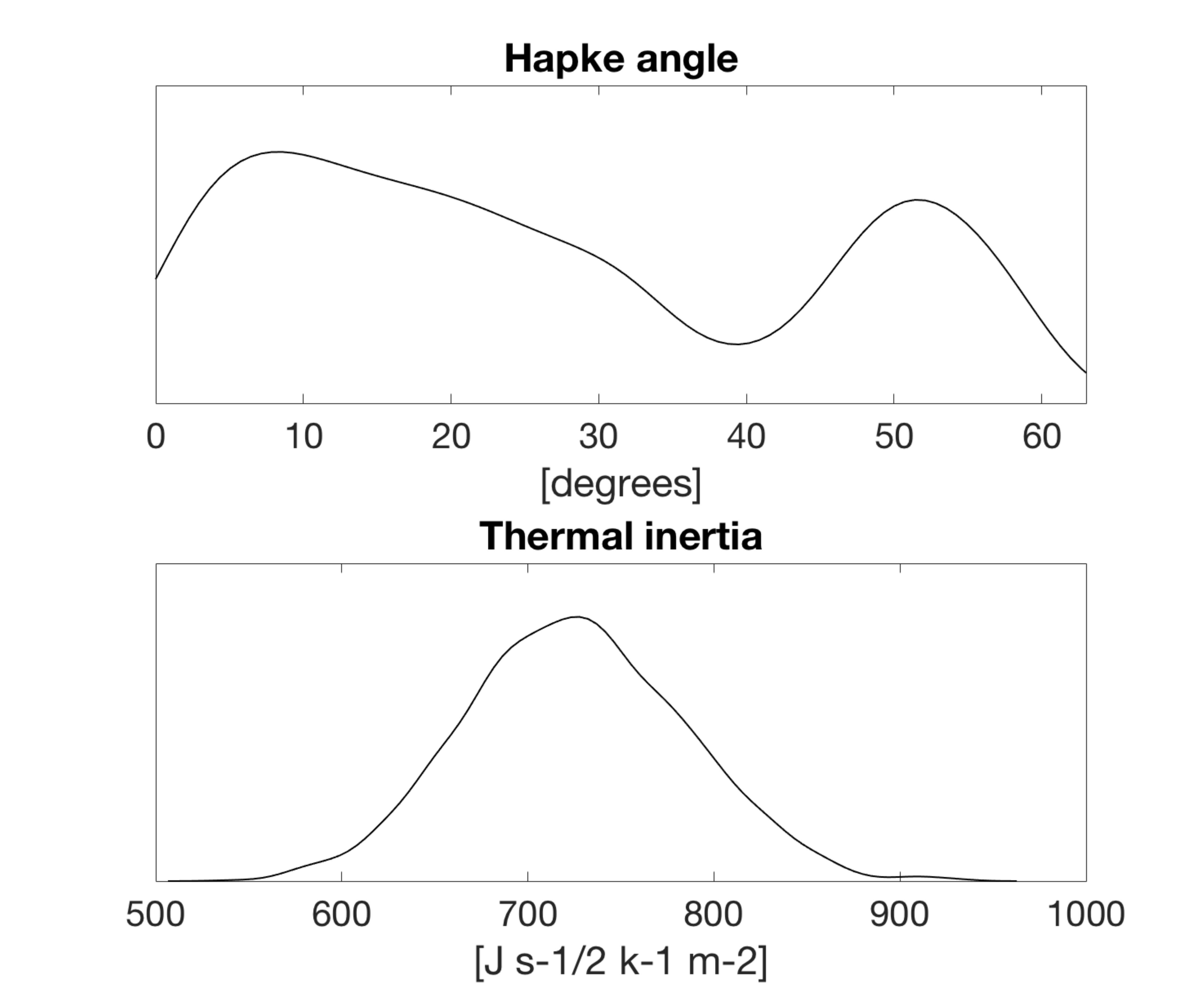}
   \includegraphics[width=\linewidth]{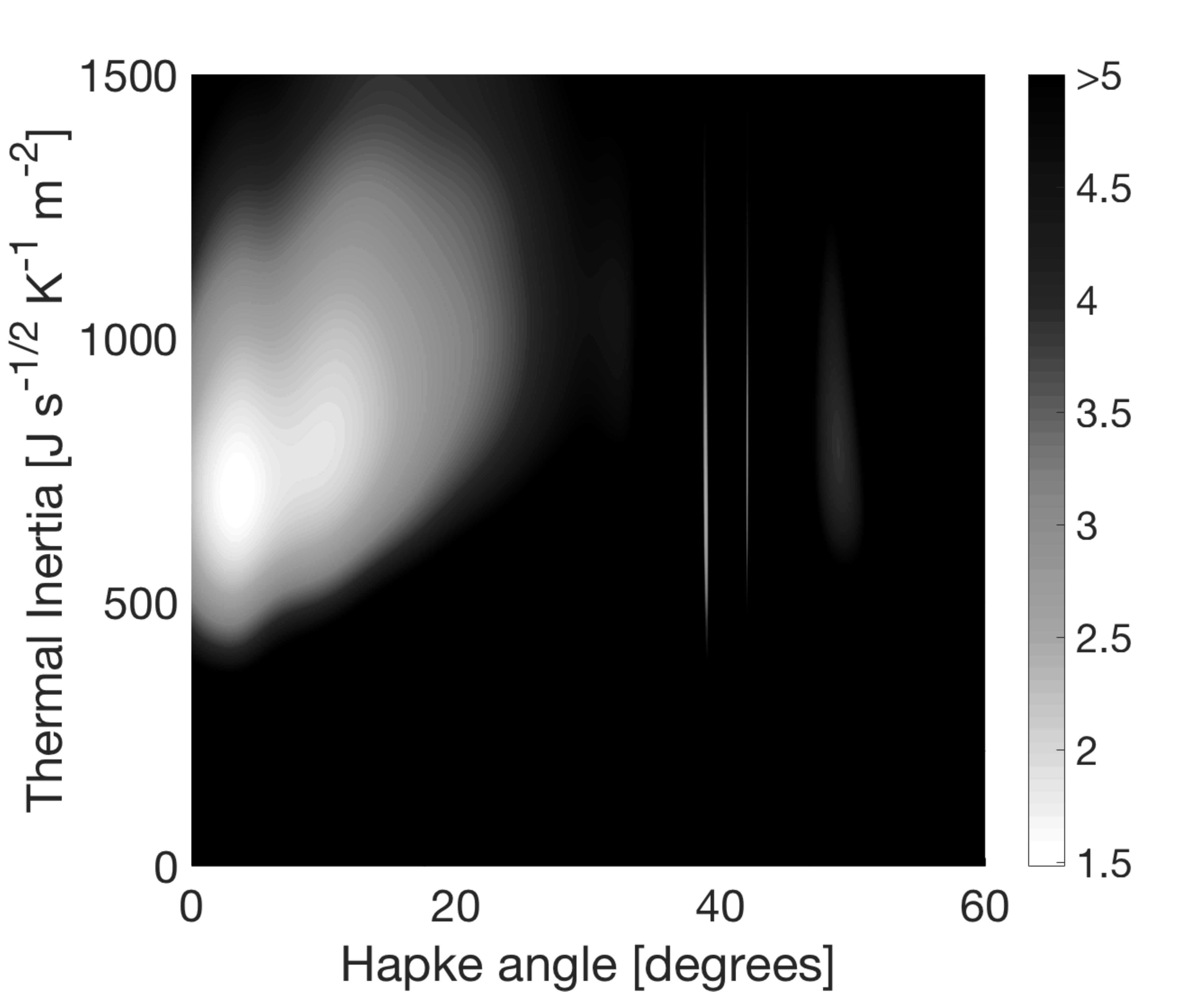}
  \caption{Top panel: posterior distributions for the Hapke angle and the average thermal inertia of the surface of Itokawa, from the Bayesian inversion of the 2-D model, using the observations in Table \ref{obs_ito}. The inversion is operated using all the fluxes and it is not informed about the Hapke angle, while the \textit{a-priori} distribution for the thermal inertia is $N(750, ~100^2)$. Bottom panel: $\chi^2_r$ phase space for  $(\bar{\theta},~\Gamma_{av.})$ parameter space. We used a selection of the available fluxes, Section \ref{ito2D}. The absolute minimum is located at around (4$^\circ$, 720 \inertiaunit).}
  \label{ito_2D}
\end{figure}

The 2-D surrogate models (inputs: surface roughness and average thermal inertia) are used as forward models in the Markov Chain Monte Carlo Bayesian inversion of the observed infrared fluxes in Table \ref{obs_ito}. The inversion of the 2-D dataset allows constraining the Hapke angle of the surface and the average thermal inertia, which is both contributed by the regolith and the rock components. The posterior distributions for the inversion of the 2-D problem are in Figure \ref{ito_2D}, top panel. We retrieve an average thermal inertia between 700 and 800 \inertiaunit, while the distribution of the Hapke angle is bi-modal.

The presence of two peaks for the roughness parameter is intriguing because of what is known about the surface of Itokawa. Analyses of Hayabusa data assessed the presence of both a smoothed terrain and a rough terrain, which have been found to correlate with low- and high- land regions respectively \citep[e.g.,][and references therein]{2006Yano,2018Susorney}. While the average thermal inertia is well constrained ($\Gamma_{av.}=721 \pm 45$ \inertiaunit), the posterior distribution for the Hapke angle is broad and does not provide any constraint on the roughness. Interestingly, however, our estimate of the Hapke angle is $26~\pm ~18^\circ$ (statistical mean and standard deviation). This is consistent both with the results from \citep{2008kitazato}, who found a photometric roughness of $26~\pm~1^\circ$ in terms of Hapke angle for the near-infrared range 0.76 - 2.25 $\mu m$, and with the average value derived for main-belt asteroids \citep{1999Mueller}.  \citet{2005MuellerT} found an average thermal inertia of the surface of Itokawa between 500 and 1000 \inertiaunit, and therefore proposed a value of 750 \inertiaunit. The Hapke angle, by contrast, remained unconstrained, and the "canonical" value for Main Belt Asteroids was assumed \citep{1999Mueller}, in terms of surface slope root mean square $\rho$ and a surface crater density $f$: $\rho=0.7$, $f=60\%$. The definition of surface slope root mean square, $\rho$, is that of \citet{1998Lagerros}:

\begin{equation}
    \rho=\sqrt{f\frac{ln(1-2S) - 2S (S-1)}{4S (S-1)}}
\end{equation}
\noindent
where $S = 0.5~(1-\cos{\gamma_c})$, with $\gamma_c$ equal to the semi-aperture angle of the crater. Values of $\rho=0.7$, $f=60\%$ correspond to a Hapke angle of about 26$^\circ$.

\begin{figure}[h!]
\centering
  \includegraphics[width=\linewidth]{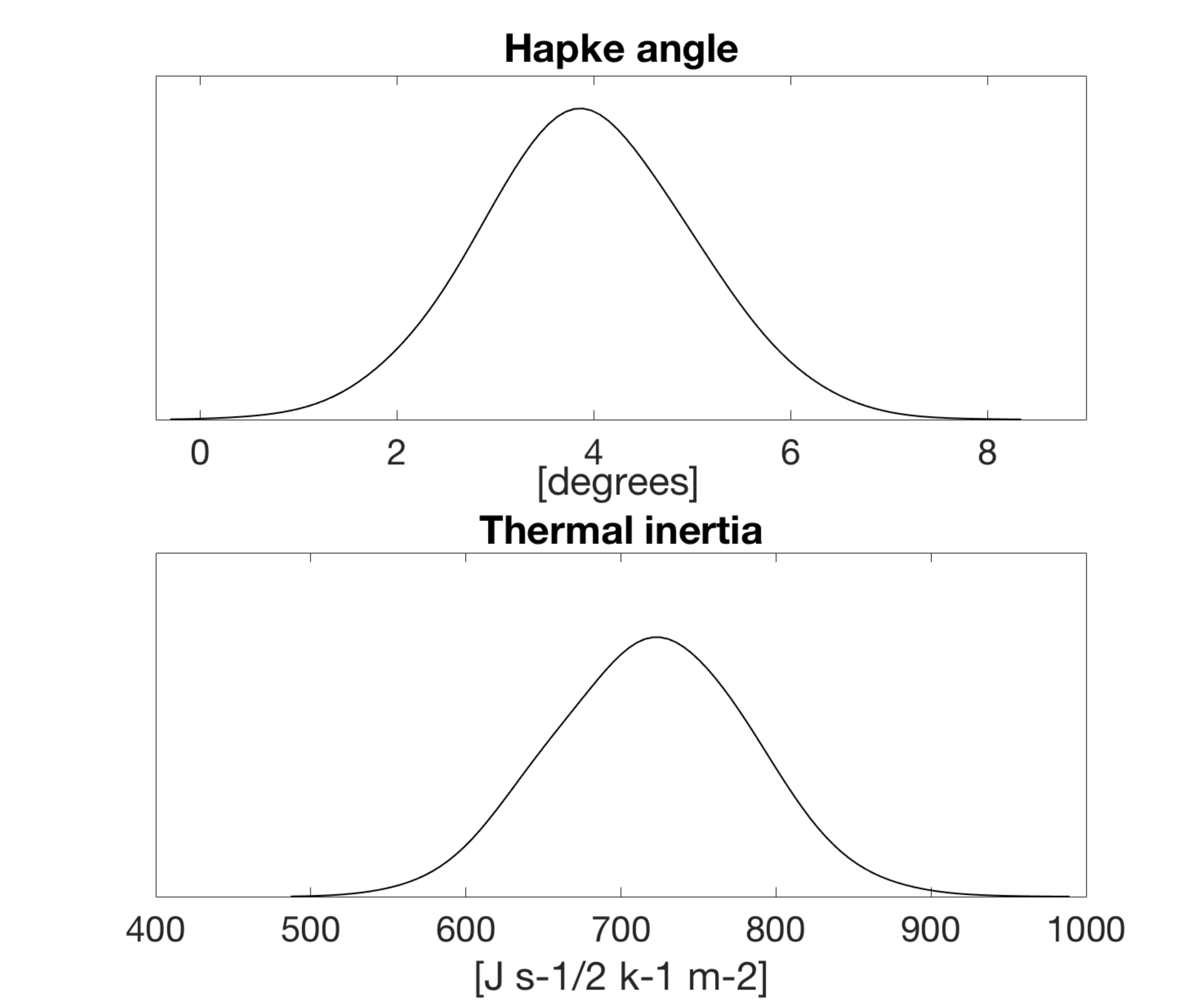}
  \caption{Bayesian inversion of the selected data, informed by the detailed $\chi^2_r$ survey. The retrieved surface properties, $\bar{\theta}$ = $4 \pm 1^\circ$ and $\Gamma_{av.}$ = $721 \pm 45$ \inertiaunit.}
  \label{ito2D_withpriori}
\end{figure}

To better investigate the bimodality, we perform a classical reduced $\chi^2$ analysis using the surrogate models in the forward mode, to explore the ($\bar{\theta}$, $\Gamma_{av.}$) phase space. The capability of the networks to generalize the responses for predictors different from the training set enables a fine mapping of the parameter space.
After a first iteration, five observations show residuals bigger than 2.5 the standard deviation of the measurement ($\sigma$-s in column 3, Table \ref{obs_ito}). The observed fluxes could be affected by unknown or/and unreported errors, as the data are an ensemble of observations performed in different conditions and at different epoch. Applying the Chauvenet criterion for data rejection, we remove the five observations (n. 4, 6, 21, 22, 23 in Table \ref{itodat}) and repeat the survey. The phase space $(\bar{\theta},~\Gamma_{av.})$ of the selected data is in Figure \ref{ito_2D}, bottom panel. The $\chi^2_r$ function shows multiple minima, at different Hapke angles, but at a consistent value of thermal inertia around 750 \inertiaunit, in agreement with previous findings \citep{2005MuellerT}. The solution (statistics of the absolute minimum at $\chi^2_r$ = 1.5) is used to inform the Bayesian inversion. The resulting posterior distributions are in Figure \ref{ito2D_withpriori}; the estimated surface properties are $\bar{\theta}$ = $4~\pm~1^\circ$ and $\Gamma_{av.}$ = $721~\pm~45$ \inertiaunit. 

\subsubsection{Bayesian inversion of the 4-D dataset}
\label{ito4D}

The 4-D scheme models the surface in terms of 4 parameters: Hapke angle, thermal inertia of the regolith, thermal inertia of the rock, rock abundance. The last three surface properties contribute to the average thermal inertia of the surface according to their weights (i.e. rock abundance). The 4-D surrogate models are used as forward models in the Markov Chain Monte Carlo inversion of the observed infrared fluxes in Table \ref{obs_ito}. The results from the inversion of the 2-D dataset inform the inversion of the 4-D model in two ways. \textit{A priori}, the Hapke angle is informed by the distribution found in Section \ref{ito2D}. \textit{A posteriori}, if the distributions of thermal inertia of the regolith ($\Gamma_{regolith}$), thermal inertia of the rock ($\Gamma_{rock}$) and rock abundance are linearly combined to give the disk-integrated thermal inertia, i.e., Equation \ref{gamma_tot}, the post-processed distribution must be consistent with the analogous result of the invertion of the 2-D dataset.

We firstly perform a fully unconstrained Bayesian inversion of all the fluxes using only the \textit{a-priori} information on the Hapke angle (derived from the inversion of the 2-D dataset).  Unfortunately, the fully unconstrained approach fails in splitting the regolith's and rock's contributions to the thermal inertia. Therefore we need to find admissible ranges for the parameters to support the inversion process. We recall some information from the retrievals of the Hayabusa spacecraft, as discussed below.  Using these constraints, we repeat the inversion and successfully retrieve the surface properties.

\begin{landscape}
\begin{table}
\centering
\begin{tabular}{|c|c|c|}
 \hline
 \textbf{Property} & \textbf{Value} & \textbf{Source} \\
 \hline
Particle Density	&	3220 $kg~m^{-3}$	&	\citet{2008Consolmagno}, average value for LL chondrites	\\
Specific Heat	&	682 $J~kg^{-1} K^{-1}$	&	\citet{2013Wach}, LL chondrite NWA 4560	\\
Particle Thermal Conductivity	&	1.5 $Wm^{-1} K^{-1}$	& \citet{2012Opeil,2017flynn}; assuming microporosity of 0.082. 	\\
Gravity	&	8.6e-5 $m s^{-2}$	&	\citet{2006Scheeres}\\
Young's Modulus	&	$7.80~10^{10}$ $Pa$ 	&	\citet{1995schultz,2013Gundlach}	\\
Poisson's Ratio	&	0.25	& \citet{1995schultz,2013Gundlach}	\\
Surface Energy	&	0.02 $J m^{-2}$	&	\citet{2013Gundlach}	\\
Emissivity	&	1	& \citet{2013Gundlach,Sakatani2018Icar..309...13S} (assumed)\\
 \hline
 \end{tabular}
 \caption{A measurement of the thermal inertia of the regolith can be converted into mean radius of the grain. Such a conversion requires the knowledge of the listed parameters. Some of these parameters -- noticeably the specific heat, $c_p$, and the particle thermal conductivity, $k$ -- are temperature dependent and have been therefore computed at a mean surface temperature of 300 K, relative to the portion of the asteroid in the field of view (the heliocentric distance is about 1 au for each observation).}
 \label{itokawa_literature}
 \end{table}
\end{landscape}
 
~\\
\textit{{A-priori} information: rock abundance.} The size distribution of blocks on the surface of Itokawa, derived from Hayabusa/AMICA images \citep{2014Mazrouei}, allows setting a lower limit for the rock abundance. The integral of the size distribution is 16\%, indicating the abundance of rock bigger of around 5 meters (the limiting resolution). This value is therefore a lower limit for the rock abundance, as the infrared flux is contributed also by smaller boulders.\\

\textit{\textit{A-priori} information: thermal inertia}. The thermal inertia of the rock ranges between the lowest thermal inertia recorded for stony meteorites -- around 700 \inertiaunit~\citep[Cold Bokkeveld, CM2 chondrite,][]{Opeil2010Icar..208..449O}, and 2500 \inertiaunit. A constraint on the thermal inertia of the regolith can be derived by exploring the relationship between regolith's average particle size and thermal conductivity. 

To interpret the fine-particulate component thermal inertia derived from this work in terms of particle size, we utilize the methodology of \citet{2013Gundlach}. This is involves comparing theoretical  values for regolith thermal conductivity at different particle sizes to thermal conductivity values derived from the predicted thermal inertia, leaving regolith porosity as a free parameter. In lieu of the particulate thermal conductivity model developed and utilized by \citet{2013Gundlach}, we instead utilize the improved model by \citet{Sakatani2017AIPA....7a5310S} along with updated model tunable parameter values from the experimental work by Ryan (2018). The model by Sakatani calculates the solid and radiative components of the thermal conductivity of particulates as a function of relevant material and environmental parameters (Table \ref{itokawa_literature}). Parameters $\zeta$ and $\xi$ are used to tune the radiative and solid conduction components of the model, respectively, in order to fit the model to experimental measurements of particulate thermal conductivity. \citet{Sakatani2017AIPA....7a5310S,Sakatani2018Icar..309...13S} provide values for particulates up to approximately 1 $mm$ in diameter. Recently, \citet{2018Ryanphd} obtained values for the parameters from conductivity experiments for particles up to 1 $cm$ in diameter. The values presented in that work can reasonably be related to particle diameter, $D$, with a power function, where $\zeta = 0.149D^{-0.3052}$ and $\xi = 0.899 D^{0.2588}$. Values for all other model parameters and their sources are provided in Table \ref{itokawa_literature}. \citet{Nakamura2011Sci...333.1113N} report that the grains retrieved from Itokawa by Hayabusa are identical to those of thermally metamorphosed LL chondrites. Thus, average material properties for LL chondrites are used where available. 

In order to directly relate the thermal inertia to particle size, a range of possibility regolith macroporosity values must be considered. A very loose, random packing of spheres has a maximum porosity of about 0.45 under terrestrial conditions where gravitational forces exceed electrostatic attractive forces between particles \citep{dullien2012porous}. Some angular basaltic samples used by \citet{Sakatani2018Icar..309...13S}, however, had porosity values up to 0.66. \citet{2014KiukiNakamura} develop a relationship between particle size and porosity on small bodies with very small surface accelerations where inter-particle forces might instead dominate. Using their model and the nominal values for silicate particles provided in their paper, and a maximum surface acceleration of 0.086 $mm s^{-2}$ for Itokawa \citep{2006Scheeres}, we estimate the maximum porosity for different particle sizes. The computed values for particles with diameters of 0.5 $mm$, 1 $mm$, 5 $mm$, 1 $cm$, and 5 $cm$ are 0.86, 0.83, 0.74, 0.7, and 0.58 (using a cleanliness ratio of unity, which is recommended for particles in space). Given the likely angular nature of the particles on Itokawa, we assume that porosities as high as 0.66 are possible for all particle size ranges. Dense random packings of spheres have a minimum porosity of about 0.36. However, we consider such a dense packing unlikely for angular particles and instead adapt a more modest minimum porosity of 0.4.

The assumption that the regolith can be treated as homogeneous is no longer valid once the particles size  exceeds the length of the diurnal skin depth \citep[chapter by Mellon et al. in][]{2008Bellbook}. For each value of thermal conductivity predicted by the \citet{Sakatani2017AIPA....7a5310S} model, a corresponding diurnal skin depth is calculated (Equation \ref{skin}). If the particle size exceeds the skin depth, these values are rejected. Although the procedure tends to lead to the rejection of larger particle sizes, this does not preclude the possibility of the existence of large particles; it simply means that the model cannot be used to confidently make interpretations within this particle size range. Figure \ref{ito_regdiam} shows the relationship between particle size, porosity and thermal conductivity. The thermal conductivity curves are dotted if the particle size exceeds the calculated skin depth; where the curves are dotted, these are expected to deviate upwards from the apparent power law trend to eventually reach a constant value for thermal conductivity that corresponds to the conductivity of solid rock. The maximum value predicted with confidence by the Sakatani model is to 0.093 $W~m^{-1}~K^{-1}$ for a porosity $\phi = 0.6$; this corresponds to a value of the thermal inertia of the regolith of about 285 \inertiaunit. We use this value to inform the inversion regarding the upper limit of thermal inertia of the regolith. 

As previously discussed, existing models for regolith thermal conductivity can only be reliably used for particle sizes less than a diurnal skin depth. As such, there potentially exist a range of particle sizes larger than the diurnal skin depth, but still small enough that their thermal signature is below that of an infinite bedrock. We choose to avoid interpretations within this range by utilizing the hard boundary for regolith thermal conductivity, but we do acknowledge that this topic needs additional research. In doing so, we make the assumption that the two extremes in thermal cycling (i.e. low amplitude thermal cycling of rock and high amplitude thermal cycling of the finest regolith) will dominate the observed signature, and that the presence of an intermediate particle size regolith within this "transition" zone would not significantly impact the end results of our interpretations. A smooth transition from rock into regolith should be expected, as the result of comminution by thermal fatigue fragmentation \citep{Delbo2014Natur.508..233D,ElMir2016LPI....47.2586E,Molaro:2011tr} and micrometeoroid bombardment \citep{Basilevsky2013P&SS...89..118B,Horz1975Moon...13..235H,Horz1997M&PS...32..179H}.\\

\begin{figure}[h!]
\centering
  \includegraphics[width=\linewidth]{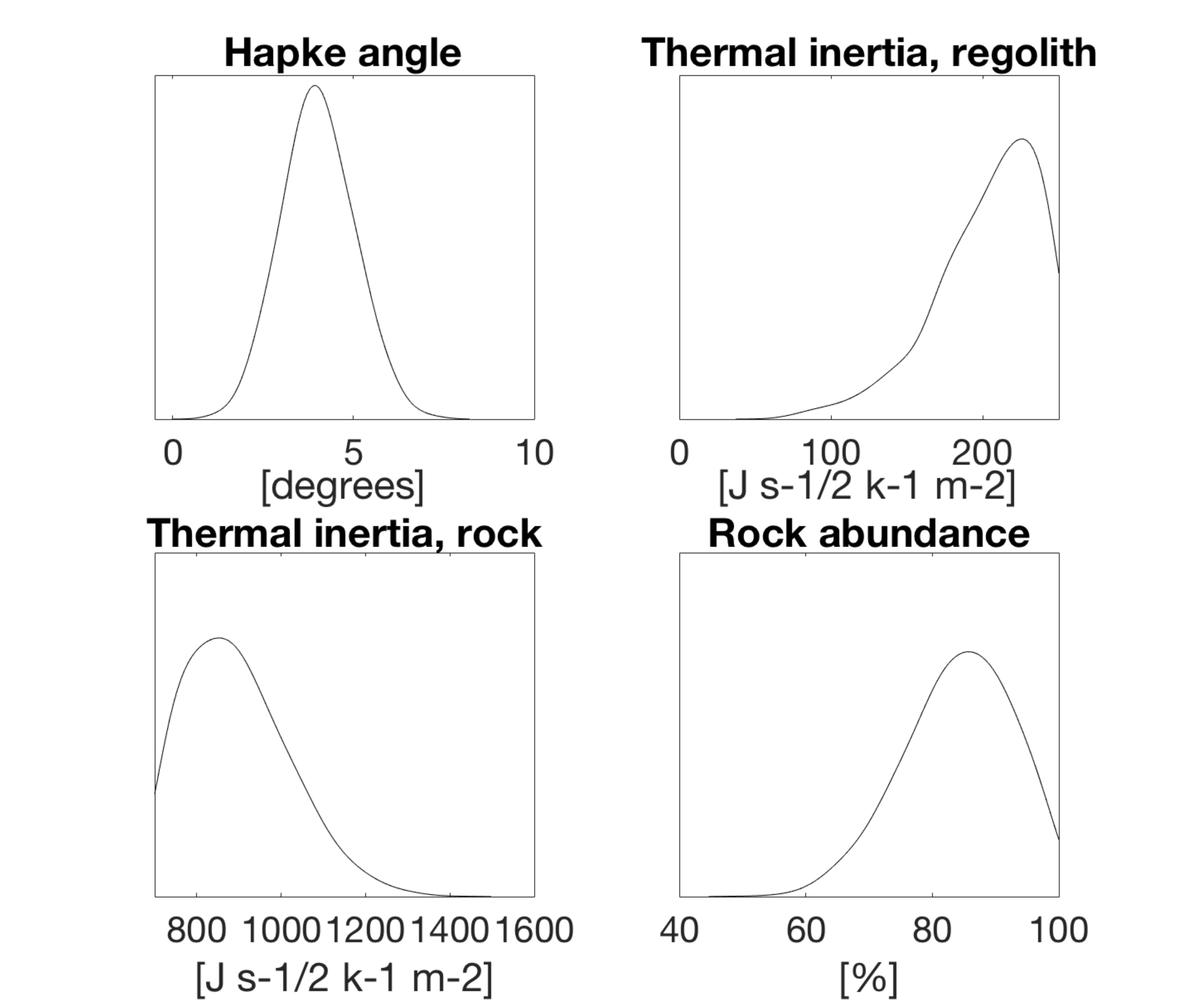}
  \caption{Posterior distribution of the parameters in the Bayesian inversion of the 4-D problem. The inversion is informed by the detailed $\chi^2_r$ survey. The retrieved surface properties are: $\bar{\theta} = 4 \pm 1^\circ$,  $\Gamma_{regolith}=203 \pm 36$ \inertiaunit, $\Gamma_{rock}=894 \pm 122 $ \inertiaunit, RA = $84~\pm 9\%$.}
  \label{ito4D_withpriori}
\end{figure}

\begin{figure}[h!]
\centering
  \includegraphics[width=\linewidth]{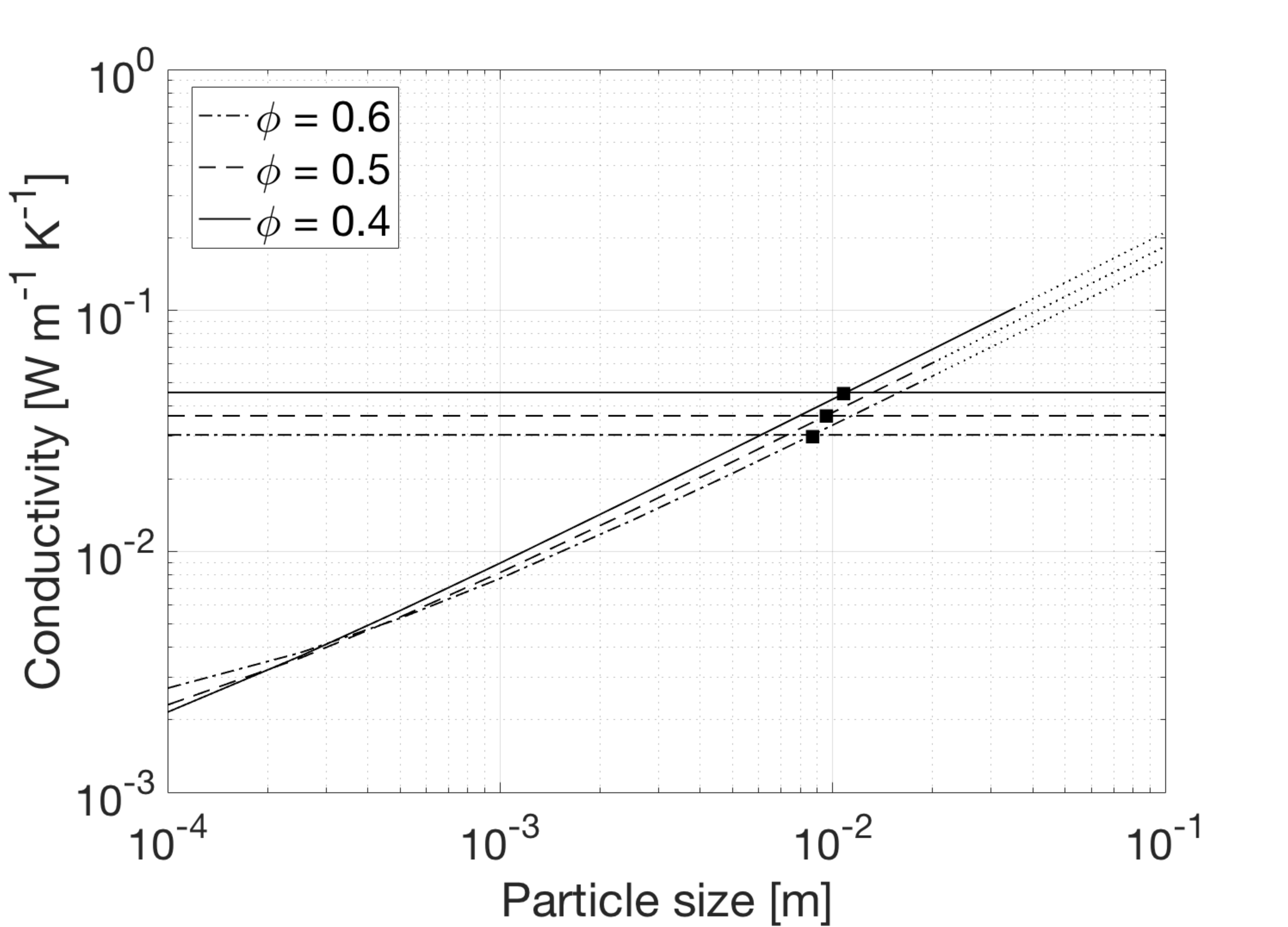}
  \caption{Thermal conductivity of regolith on Itokawa as a function of mean particle diameter, for three different values of porosity. The curves are dotted for values of particle size exceeding the calculated skin depth. The horizontal lines are values of conductivity for a nominal thermal inertia for the regolith equal to $203$ \inertiaunit, at different values of porosity. Solutions (intersections between thermal conductivity curves and horizonatal lines) exist for $\phi = 0.6$, $\phi = 0.5$ and $\phi = 0.4$. The retrieved grain size is in line with in-situ findings by \citep{Yano2006Sci...312.1350Y,2008kitazato}. }
  \label{ito_regdiam}
\end{figure}

Using the above constrains, we perform a detailed $\chi^2_r$ survey on all the observation but those rejected in the inversion the 2-D simulation. The \textit{a-priori} information available from literature allows focusing the search on admissible sub-regions of the $\chi^2_r$ hyperspace while looking for the absolute minimum. The solution (statistics of the absolute minimum at $\chi^2_r$ = 1.6) is used to inform the Bayesian inversion of the 4-D model. Figure \ref{ito4D_withpriori} shows the \textit{posterior} distributions for the parameters; the best estimates have statistics: $\bar{\theta} = 4 \pm 1^\circ$,  $\Gamma_{regolith}=203 \pm 36$ \inertiaunit, $\Gamma_{rock}=894 \pm 122 $ \inertiaunit, RA = $84~\pm 9\%$.

\subsubsection{Discussion}

\begin{figure}[h!]
\centering
  \includegraphics[width=\linewidth]{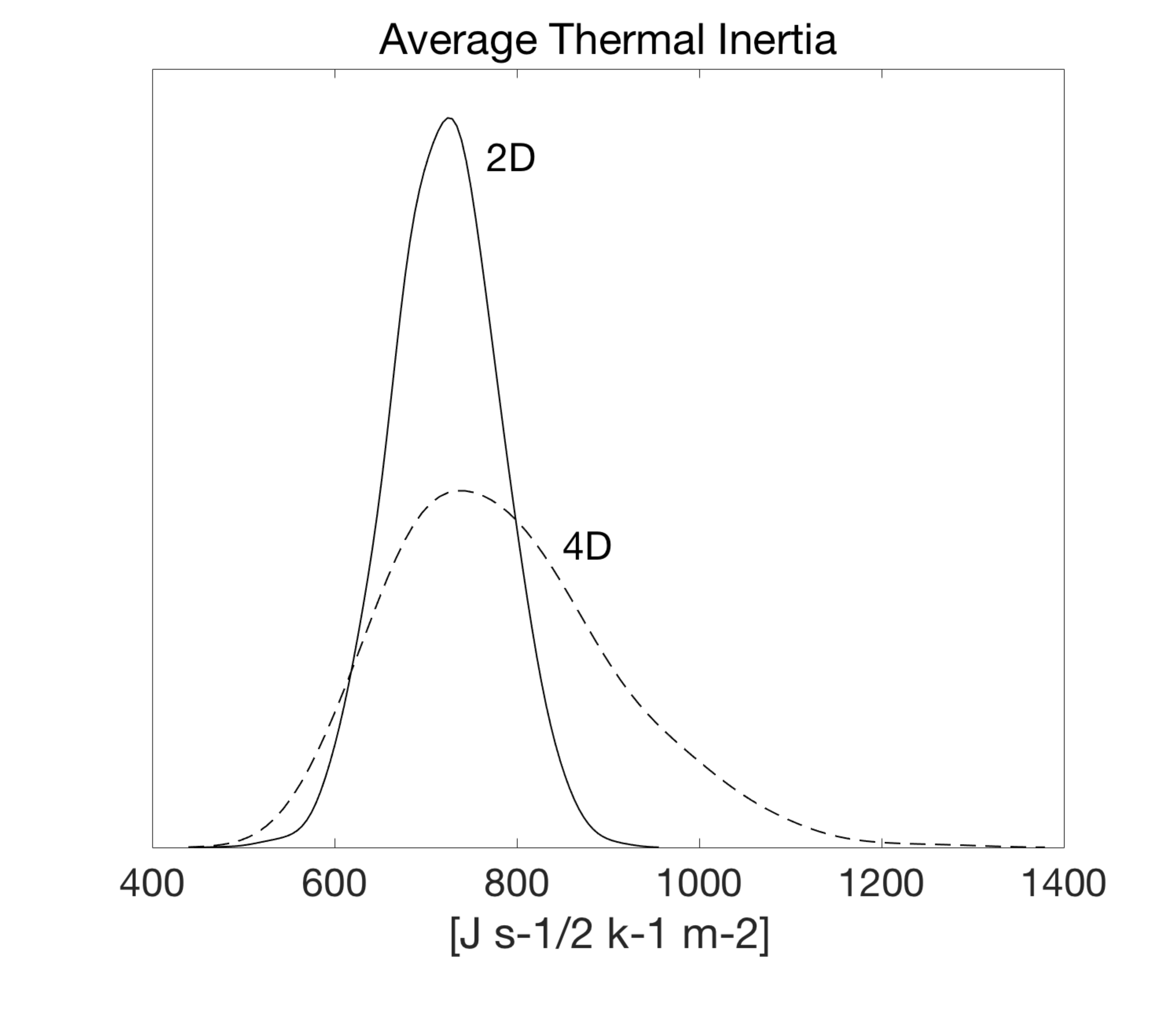}
  \caption{The posterior distributions of thermal inertias (regolith and rock) are combined according to the posterior distribution the rock abundance (bottom right panel in Figure \ref{ito4D_withpriori}); the post-processed average thermal inertia of the surface (dashed curve) has a statistics equal to $\Gamma_{av.,~4D} = 785 \pm 119$ \inertiaunit. The solid curve is the posterior of the same quantity from the inversion of the 2-D dataset (bottom panel of Figure \ref{ito2D_withpriori}, $\Gamma_{av.,~2D}$ = $721 \pm 45$ \inertiaunit). The two curves peak at the same value of $\Gamma_{av.}$ and their statistics overlap at less than 1--$\sigma$.  The 4-D solution to average thermal inertia, however, has a broader posterior distribution with respect to the 2-D solution; we attribute this degradation to error propagation.}
  \label{ito4D_comp}
\end{figure}

We compare the results of the inversion of the 4-D dataset to those of the 2-D dataset as well as to in-situ measurement of regolith size. The average thermal inertia of the surface is re-constructed as the combination of the posterior distributions of the thermal inertias (regolith and rock) and the rock abundance. Figure \ref{ito4D_comp} shows the average thermal inertia from the inversion of the 2-D dataset (solid line) versus the post-processed average thermal inertia from the inversion of the 4-D dataset (dashed line). Both curves suggest a value of thermal inertia around $750$ \inertiaunit ~($\Gamma_{av.,~2D}$ = $721 ~\pm~45$ \inertiaunit~and $\Gamma_{av.,~4D} = 785~\pm~ 119$ \inertiaunit, for the 2-D case and 4-D case respectively). The two curves peak at approximately the same value of $\Gamma_{av.}$ and the estimated values overlap at less than 1--$\sigma$. Both results are consistent with previous findings \citep{2005MuellerT}. The 4-D solution to average thermal inertia, however, has a broader posterior distribution with respect to the 2-D solution; we attribute this degradation to error propagation.

The value of thermal inertia of the regolith derived in Section \ref{ito4D} is converted to thermal conductivity, using the mean density and specific heat of LL chrondrites, for different porosity values (horizontal lines in Figure \ref{ito_regdiam}). Where corresponding porosity lines intersect, a prediction for particle size is made, following the method by \citep{2013Gundlach}. Solutions exist for porosity equal to 0.4, 0.5 and 0.6 (squared dots in Figure \ref{ito_regdiam}). After error propagation, the mean particle diameter is found to be: $9^{+6}_{-4}$ $mm$ ($\phi = 0.4$); $10^{+7}_{-4}$ $mm$ ($\phi = 0.5$); $11^{+7}_{-5}$ $mm$ ($\phi = 0.6$). These results are in agreement with in-situ findings by Hayabusa \citep{Yano2006Sci...312.1350Y,2008kitazato}.

The value of thermal inertia of the rock derived at the end of Section \ref{ito4D} is converted to thermal conductivity, again using the mean density and specific heat of LL chondrites, for a microporosity of 0.082. After error propagation, we retrieve a value of $0.4\pm0.1~W~m^{-1}~K^{-1}$. This value is substantially lower than the thermal conductivity of coherent LL chondritic rocks, i.e., around 1.5 $~W~m^{-1}~K^{-1}$ \citep[Figure 2 in][]{Opeil2012M&PS...47..319O}; we therefore conclude that the rocks on Itokawa could be fractured.

Finally, the minimum $\chi^2_r$ for the 4-D model is 1.6. Within 1--$\sigma$ ($\sigma_{\chi^2_r} = 0.3$), this value is the same of the 2-D model and, in this respect, the two models are equivalent. In addition to the 2-D model, however, the 4-D model is able to split the contribution from the regolith and the rock to the average thermal inertia, and therefore is able to inform us about the rock abundance on the surface of the asteroid without any assumption on the thermal inertia of the components (besides the discussed upper and lower limits). 

\section{Conclusion}
\label{Conclusion}

Infrared fluxes emitted from asteroids carry information about the properties of the surface. The customary approach to the inverse problem (i.e., backing up surface properties from observed infrared fluxes) is to run thermophysical simulations, in which the asteroid thermal properties are mapped into an infrared flux. The flux is then compared to the observed data following $\chi^2$ minimization (see Section \ref{method}).  Using this approach, past studies \citep[e.g.,][]{2002Harris,2005MuellerT, 2006Emery,2014Emery,Delbo2009P&SS...57..259D,2015Delbo,2014Rozitis} have been successful in retrieving size, albedo and/or average thermal inertia of asteroids, but others important properties, such as surface roughness and rock abundance, remained elusive and unconstrained. Many difficulties arise when multiple surface properties are included in the game. The number of simulations needed to densely populate the parameter space grows exponentially with the dimension \textit{N} of the input array. The objective function to be minimized (e.g., $\chi^2$ function in Equation \ref{chi2eq}), may show local minima or saddle points also for relatively low values of \textit{N}, as shown in Figure \ref{ito_2D} for the interpretation of disk-resolved thermal infrared data of asteroid (25143) Itokawa using the 2-D dataset. In order to split the contribution of regolith and rock to thermal inertia (e.g., Figure \ref{ito4D_withpriori} for Itokawa), we want to increase the dimensionality of the input array to \textit{N} = 4. For this task, the use of the TPM code results in a coarse mapping of the parameter space unless we employ significant computational resources, or we generalize the sparse simulations, i.e., by training neural networks.

Our new methodology relies on the concept of (thermal) surrogate model, which is a neural network representation of a thermophysical scheme. The ``parent" model is used in forward mode to populate a dataset of thermal simulations, which are then used to train, validate and test the networks (Section \ref{ourmethod}). The surrogate model assimilates the model and predicts answers at a known level of accuracy (with respect to the testing examples) and in an extremely low computational time. These properties enable detailed mapping of the parameter space, as it was not possible before. Because of the above, the surrogate models are "statistically invertible"; they can be used to sample the (unknown) posterior distributions of the input parameters which is associated to an observed output. In our study, we adopted a Bayesian approach to inversion, using a Markov Chain Monte Carlo technique. A MCMC algorithm runs ten of thousands of simulations, inputting guesses of the solution associated with an observed set of data. The output is then compared to the observed data, and the residual is used to compute the probability that the guessed set of parameters is the optimal solution. 

The developed methodology has been tested on two different cases. Firstly, we simulated a remote sensing problem for asteroid (101955) Bennu, the target of the NASA OSIRIS-REx mission, of which the infrared camera on-board the spacecraft \citep[OTES,][]{2018ChristensenOTES} will retrieve the emitted infrared fluxes. We used the TPM model \citep{2015Delbo} to generate a grid of thermal simulation, and then we used it to train a set of neural networks. The peculiar geometry of the observations (which allows observing also the night side of the asteroid), the wide range of wavelengths that will be observed, as well as the expected high quality of the observations (characterized by the signal-to-noise-ratio) allow retrieving the surface properties of the asteroid - surface roughness, thermal inertia of the regolith and rock, and relative rock abundance - at a high accuracy, without suggesting any \textit{a-priori} information to the Bayesian algorithm (Section \ref{Bennu}). The accuracy of the method is expected to further increase if additional information about surface properties provided by other OSIRIS-REx instruments (e.g., OVIRS, OCAMS) are recalled. 

The use of uninformative \textit{a-priori} distributions, however, is not always possible. When data are partial (e.g., ground-based observations, which cannot image the night-side of the asteroid), and/or we lack information at long wavelengths, the use of well-educated guesses about surface properties of the asteroid must be employed. This necessity clearly emerged during our second exercise, in which we inverted observed infrared flux of asteroid (25143) Itokawa, the target of the Hayabusa mission (Section \ref{Itokawa}). Data from the spacecraft have been coupled with models to firstly inform the Bayesian inversion and therefore verify the consistency of the retrieved quantities. Following the proposed procedure (creation of the training dataset, training of the surrogate model, Bayesian inversion), we constrained the surface properties of the asteroid. In Section \ref{ito2D}, we discuss an apparent bimodality in the posterior distribution of the surface roughness, which qualitatively matches with in-situ observations of the presence of both smoothed and rough terrains on the asteroid \citep[e.g.,][and reference therein.]{2006Yano,2018Susorney}. In Section \ref{ito4D}, we retrieve the thermal inertia of the regolith and the rock components, and relative rock abundance. The rock abundance on the surface (i.e., abundance of material whose size is larger than the diurnal skin depth) is found to be around 85\%. The thermal inertia of the regolith suggests a mean particle size around 10 $mm$ in diameter, in agreement with previous findings \citep{Yano2006Sci...312.1350Y,2008kitazato}. The value for the rock indicates a low thermal conductivity (if compared to LL chondrites), which could suggest the presence of fractured rocks. If the retrieved parameters are combined (Equation \ref{gamma_tot}), we get an average value of thermal inertia equal to $785 \pm 119$ \inertiaunit, consistent with previous findings by \citet{2005MuellerT}, who suggested a value around 750 \inertiaunit~(Figure \ref{ito4D_comp}).   

The present study is optimized towards the interpretation of infrared fluxes by spacecraft mission operating in proximity of airless bodies. In such scenario, the asteroid properties (e.g., albedo and size) and morphology are well characterized by in-situ measurements, and the infrared response of each facet of the shape model can be characterized by means of an associated surrogate model, leading to a global mapping of the surface properties. On the other hand, the interpretation of disk-integrated data could be affected by heterogeneity in rock abundance as well as presence of large boulders of the surface; the inversion of the surrogate models informs about the co-presence of regolith and boulders on the surface, while it does not resolve their relative distribution over the surface. Additionally, when shape models have non-negligible uncertainty in the size, the optimization of size or albedo as free parameters should be included. We will discuss how to overcome the application limit of our method to disk-resolved data in a future work.

We foresee the use of the proposed methodology to a variety of infrared remote sensing problem. Immediate applications are the analysis of (101955) Bennu by means of forthcoming data from the OSIRIS-REx spacecraft (this case has been already modeled in Section \ref{Bennu}) and of (162173) Ryugu, the target of the JAXA's Hayabusa 2 mission \citep{2013Kuninaka}. Both spacecrafts will acquire infrared fluxes (by using an infrared spectrometer and an infrared camera, respectively) in close proximity of the asteroid, therefore providing infrared fluxes also of the night-side. Our method may be used to characterize the fine-scale composition, which is important for shedding lights on the collisional history, as well as the surface evolution, of the target asteroids. In case of application to spacecraft data collected during an extended period of time, the change in heliocentric distance can be taken into account while training the networks. During the detailed survey for (101955) Bennu, however, the heliocentric distance of the asteroid will vary between 1.0 and 1.25 au, resulting in a change of the surface temperature of at most 10 K. We consider this a very small effect probably obscured by other sources of errors. In any case, we expect no difference in the accuracy of the determination of the surface physical parameters as long as the thermo-physical parameters are temperature independent. We will study temperature depended thermal inertia in a future work. Finally, the finding that rocks on Itokawa could be fractured is also relevant to the interpretation of future infrared observations of (101195) Bennu and (162173) Ryugu. Given the small size of these targets and their high rock abundance, differences in thermal inertia could be representative of more or less fractured rocks, rather than indicating the presence of regolith material (i.e., pebbles with size smaller than the diurnal skin depth).

The concept of thermal surrogate model is not limited to asteroid studies, but it can be applied whenever a thermal model is available. Measurements of comet's surface temperature \citep[e.g., by ROSETTA/VIRTIS at comet 67P/C-G,][]{1999VIRTIS} can also be inverted in their regolith and rock contributions, providing insights in the processes of evolution of cometary surfaces. Improved rock abundance maps on Mars and the Moon can be generated, thanks to the large availability of thermal data from previous and current missions. In future, infrared fluxes of Jupiter's moon Europa and Ganymede will be provided by the NASA's Europa Clipper and the ESA's JUICE missions; a surrogate thermal model of the outer ice shell behaviour would allow the characterization of the surface properties, informing future landing mission about sampleability at the diurnal skin depth level. 

\section*{Acknowledgements}
\label{Acknowledgements}
MD and AR acknowledge support from the Academies of Excellence on Complex Systems and Space, Environment, Risk and Resilience of the Initiative d'EXcellence ”Joint, Excellent, and Dynamic Initiative” (IDEX JEDI) of the Universit\'e C\^ote d'Azur, as well as the  Centre National d'Etudes Spatiales (CNES). The authors thank the anonymous referees for the precious comments and suggestions that improved this manuscript.

\section{References}
\bibliographystyle{elsarticle-harv}
\bibliography{main}

\end{document}